\documentclass[showpacs,preprint,amsmath,amssymb,floatfix]{revtex4}

\pdfoutput=1

\usepackage{graphicx}

\usepackage{dcolumn}           
\usepackage{bm}                

\begin{document}
\title{Role of chain stiffness on the conformation of single polyelectrolytes 
in salt solutions}
\author{Yu-Fu Wei} 
\author{Pai-Yi Hsiao} 
\email[Corresponding author. E-mail: ]{pyhsiao@ess.nthu.edu.tw}
\affiliation{Department of Engineering and System Science, 
National Tsing Hua University, Hsinchu, Taiwan 300, R.O.C.}
\date{\today} 

\begin{abstract} 
Conformation of single polyelectrolytes in tetravalent salt solutions 
is investigated under the framework of a coarse-grained model, using 
Langevin dynamics simulations.  
The chain size, studied by the radius of gyration, shows three different
variational behaviors with salt concentration, depending on the chain stiffness.
According to the size variations, polyelectrolytes of fixed chain length 
are classified into three categories: 
(1) flexible chain, for which the variation shows a curve similar to
a tilted letter L; 
(2) semiflexible chain, whose curve looks resemble of the letter U;
(3) rigid chain, for which the curve is a straight line.
The worm-like chain model with persistence length predicted by 
the Odijk-Skolnick-Fixman theory is found to be able to qualitatively
describe the end-to-end distance at low salt concentration,
not only for semiflexible and rigid chains but also for flexible chain.
In a low-salt region, a flexible polyelectrolyte extends more significantly  
than a semiflexible chain, in reference of the size of their uncharged counterparts, 
and in a high-salt region, regardless of chain stiffness, a chain attains 
a dimension comparable to that of its neutral polymer.
The chain stiffness influences both the local and the global chain structures. 
A flexible chain exhibits a zigzagged local structure in the 
presence of salt ions and the condensed structure is a disordered, random globule.
A semiflexible chain is locally smooth, and  
the condensed structure is orderly packed, taking a form such as hairpin or toroid.
Moreover, the chain stiffness can also affect the nature of the coil-globule transition.
The transition is occurred in a discrete manner for semiflexible chain,
whereas in a continuous way for flexible chain.
This discrete feature is happened not only at low salt concentration
when a semiflexible chain is collapsed, but also at high salt concentration
when the collapsed chain is reexpanded.  
At the end, the effects of chain stiffness and salt concentration on 
the conformation of single polyelectrolytes are summarized in a schematic state diagram.  
\end{abstract} 


\maketitle

\section{Introduction}
There is a resurgent interest in studying the properties of polyelectrolytes 
in multivalent salt solutions because such system reveals many fascinating 
phenomena.
One vital example is the DNA condensation, in which DNA, a negatively-charged 
polyelectrolyte, undergoes a dramatic condensation from an extended structure to 
a compact, highly-ordered structure while multivalent salt is added to the 
solution~\cite{gosule76,arscott90,bloomfield96,yoshikawa02b}. 
The condensation induced by multivalent salt is not a privilege reserved only for DNA 
but a common feature of polyelectrolytes~\cite{delacruz95,raspaud98,maurstad03}.
Usually trivalent salts or charged molecules of higher valence are demanded 
to induce the condensation~\cite{widom80,bloomfield96}.  
Experiments have shown that the morphology of the condensate depends strongly on the chain 
bending rigidity which balances the attractive collapsing force 
at a solvent quality. 
Flexible polymers generally collapse to disordered 
globules, and stiff ones collapse to ordered structures such as toroid or 
folded chain~\cite{arscott90,plum90,maurstad03}.
The influence of chain stiffness on chain morphology has been investigated 
by simulations, including the condensation of polyelectrolyte induced by 
multivalent salt~\cite{khan05} or by increasing Coulomb strength 
parameter~\cite{ou05} and the collapse of neutral polymer 
driven by temperature decrease~\cite{noguchi98,ivanov00}. 
The results confirmed that chain rigidity plays a decisive role in determination of
the condensed structure.

A phenomenon which is less well-known appears when an excess of multivalent salt 
is presented in the solution: The salt-induced condensation redissolves into the solution 
and the system returns to a homogeneous phase~\cite{delacruz95,raspaud98,kruyt49}.
This phenomenon is called ``reentrant condensation''~\cite{nguyen00} and 
has been discovered since seventy-five years ago in the study of the
coacervation of colloids by oppositely charged ions~\cite{bungenbergdejong31}.
It occurs for flexible chains, for example, polystyrene sulphonate~\cite{delacruz95},
and also for stiff chains, for example, DNA~\cite{raspaud98}. 
The persistence length in the latter example is two order of magnitude 
larger than in the previous.
As the condensation takes place, the bare chain charge is almost 
neutralized by the condensed counterions dissociated from the multivalent 
salt~\cite{wilson79,yamasaki_y01}.
These counterions not only screens out the long-range Coulomb repulsion 
but also creates a short-range attraction due to correlated fluctuation, 
resulting in the condensation~\cite{oosawa71,rouzina96}.
On the other hand, for the redissolution of polyelectrolytes  
at high salt concentration, no comprehensive explanation is well established.
Recently, two theories were proposed to explain the phenomenon but 
verification is still ongoing.  
The first theory suggested that the redissolution is a consequence of 
charge inversion occurring when polyelectrolyte binds so many counterions 
that its net charge alters sign~\cite{nguyen00,grosberg02}, whereas 
the second one argued that it has an origin of the screening of the short-range 
attraction~\cite{solis00,solis01}.  The second theory, furthermore, predicted 
that Bjerrum association plays an important role on the organization of 
polyelectrolyte-ion complexes~\cite{solis02}, and therefore, charge reversal is 
not always happened.

The folding transition of a polymer between an elongated state and a compact 
state is called ``coil-globule transition''~\cite{lifshitz78}.
Theorists predicted that for semiflexible chains, such transition is first order, 
using mean-field theory~\cite{lifshitz78,post82,ghosh02}. 
The prediction has been confirmed experimentally~\cite{yoshikawa95,yoshikawa96}.
Despite continuous for the whole ensemble, the coil-globule transition 
is discrete at the level of single chain
for polymers with sufficient stiffness~\cite{yoshikawa96}. 
In this case, the transition is similar to the disordered-ordered transition between 
a gas-like and a crystal-like state. 
On the other hand, for single polymers with small stiffness, 
the transition is continuous, resembling to the disordered-disordered transition 
between a gas-like and a liquid-like state.
This topic has been recently studied in simulations~\cite{ivanov00,noguchi98,ou05,khan05}. 
The results showed that the discrete transition can be induced by adding condensing 
agent~\cite{khan05} or by decreasing temperature~\cite{noguchi98,ivanov00},
while chains are stiff. 
However, the discussions were mainly focused on the chain condensation 
occurring in low temperature region or in low salt regime.
Since an excess of multivalent salt can lead condensed chains reentering into 
the solution, it becomes very relevant to know whether the reentering transition 
behaves in a discrete manner or not. 
To our knowledge, there is no simulation nor experiment, hitherto, 
discussing the conformational transition on the chain reentrance
 occurring at high salt concentrations.  
To get insight of it, we investigate in this paper the behavior of single 
polyelectrolytes with addition of multivalent salt by means of computer simulation. 
The salt covers a broad range of concentration so that single chains show, 
in turn, the condensation and the reentrance transitions in a microscopic way.  
Our study, therefore, offers a good opportunity to investigate thoroughly 
the coil-globule transition happened at low and at high salt concentrations.
  
There have been many simulations devoting to the study of the behavior of
polyelectrolytes in salt-free 
solutions~\cite{stevens93,stevens95,winkler98,chu99,micka99,stevens01,chang02,liu02,pais02,deserno03,limbach03,ou05}.
Only recently, the study with addition of salt becomes numerically feasible
because of the progress of computing power. 
However, the salt valence studied in the literatures was usually monovalent or 
divalent~\cite{stevens98,khan99,deserno01,liu03}. 
For the study with high salt valence, the focus was mainly on the 
condensation of polyelectrolyte and the salt concentration was 
not high enough to investigate the chain reentrance into a 
solution~\cite{dias03,sarraguca03,khan05,klos05a}.
One indirect way to discuss DNA condensation and redissolution is 
to measure the effective interaction between two immobile 
rigid polyelectrolytes in the environment of multivalent 
salt~\cite{deserno03,allahyarov05}.
Recently,  progress has been made in a series of study~\cite{hsiao06a,hsiao06b,hsiao06c}: 
By carefully choosing the simulation parameters, we were able to investigate 
salt-induced condensation and redissolution thoroughly 
in a direct way. 
We found that the dimension of flexible polyelectrolytes undergoes subsequently 
two continuous transitions upon addition of multivalent salt, 
the collapsing transition and the swelling transition, which are respectively 
the microscopic representation of the condensation and the redissolution of polyelectrolyte. 
We demonstrated that the excluded volume of ion plays an imperative role on the
chain properties and should be inevitably incorporated in a 
theoretical analysis~\cite{hsiao06a,hsiao06b}.
The study of the potential of mean force showed that like-charge attraction between chains 
happens only when salt concentration is intermediate around \textit{the equivalence point} 
(see Sec.~\ref{Sec:Rg} for definition) and the size ratio of ion to monomer lies within 
a window around unity~\cite{hsiao06b}.  
Moreover, it was shown that only salt valence greater than two can induce apparently 
the two structural transitions~\cite{hsiao06c}, which is 
consistent with experimental observations~\cite{gosule76,wilson79,ma94,raspaud98}. 
Nonetheless, the effect of chain stiffness was not discussed in these studies. 
Since semiflexible polyelectrolytes exhibit several intriguing phenomena, not being 
seen in flexible polymers, we focus in this work how chain stiffness influences  
chain morphology.

Toroid structure is presumed to be the ground state of a compacted 
DNA~\cite{bloomfield96,maurstad05}.  
However, folded-rod structure and a number of other morphology, 
such as racquet, are also often observed in 
experiments~\cite{chattoraj78,martin00,yoshikawa02}. 
The long lifetime of some of these structures indicates that DNA has 
complicated energy landscape.  
Numerical studies showed that the energy levels of toroid and rodlike 
structure are comparable~\cite{noguchi98,stevens01}.
These two structures are surrounded by many metastable states and
hindered by high energy barriers.
Debate continues until now in deciding which structure is the real 
ground state of a condensed DNA and under what condition it applies. 
One of the important factors which affects the chain morphology
is the chain length. 
It was observed that longer chains display higher appearance percentage 
for toroids than for rods~\cite{marquet87,arscott90,bloomfield91}. 
Also, the degree of ion condensation on a chain was found to increase with chain 
length~\cite{cheng02,sarraguca06}. 
Recently, chain length effect on the state diagram of a single chain 
has been discussed~\cite{stukan03}.  
Limited to today's computational power, time period able to be studied by
molecular simulations is very short. 
It is, thus, difficult to investigate properly the evolution of 
chain morphology, because of the long lifetime of some of the chain structures. 
Therefore, many authors found that the formation of a 
toroidal or a rod-like structure pertains to the initial chain 
configuration used in simulations~\cite{noguchi98,stevens01,ou05}.
In this article, we study the morphology of a single polyelectrolyte  
with fixed chain length.  
For the chain stiffness in which the morphological evolution is slow,
several simulations are performed, starting with different initial configurations, 
to make sampling more complete.
By this way, we calculate static properties.
The rest of the article is organized as follows.
The simulation model and the method used in this study are described in 
Sec.~\ref{Sec:Model_Method}. 
The results and discussions are given in Sec.~\ref{Sec:Result_Discussion}.
The topics include 
the radius of gyration (Sec.~\ref{Sec:Rg}),
the degree of chain swelling (Sec.~\ref{Sec:cRg}), 
the end-to-end distance (Sec.~\ref{Sec:Re}),
the probability density distribution (Sec.~\ref{Sec:pdd}), 
and the dynamics of chain conformation (Sec.~\ref{Sec:x-t}).
Simulation snapshots are presented in Sec.~\ref{Sec:snapshot}.
We summarize the conformation of a single polyelectrolyte 
in a state diagram in Sec.~\ref{Sec:state_diagram}  
and give our conclusions in Sec.~\ref{Sec:Conclusion}.

\section{Model and simulation method}
\label{Sec:Model_Method} 
Our system consists of a single polyelectrolyte
modeled by a negatively charged bead-spring chain 
and spherical counterions dissociated from the chain.
There are 48 beads (monomers) on the chain.
Each of the beads carries a $-e$ charge and
each of the counterions carries a $+e$ charge.
Tetravalent salt is added into the system.
It is dissociated into tetravalent cations (counterions) 
and monovalent anions (coions), modeled by charged spheres.
Solvent molecules are not incorporated explicitly in the simulations. 
We suppose them forming a medium of constant dielectric constant $\epsilon_r$.
The collision between particles and solvent molecules is considered, 
and the effect is described by Langevin equation.
The system is placed in a cubic box and periodic boundary condition is applied.

Four types of interaction are involved in the simulations.
The first one is the excluded-volume interaction, which is applied for 
all the particles, including monomers, counterions and coions.
It is modeled by a purely repulsive Lennard-Jones (LJ) potential
\begin{equation}
U_{ex}(r)=\left\{\begin{array}{ll}
                              4\varepsilon_{LJ}
                               \left[
                                  \left(\sigma/r\right)^{12}-
                                  \left(\sigma/r\right)^6
                               \right]
                               +\varepsilon_{LJ} & \mbox{for } r \leq 2^{1/6}\sigma\\
                               0 & \mbox{for } r > 2^{1/6}\sigma
                             \end{array}\right.
\end{equation}
where $\varepsilon_{LJ}$ is the interaction strength and
$\sigma$ is the diameter of a particle. 
The second interaction is the Coulomb interaction
$ U_{coul}(r)=Z_i Z_j e^2/(4\pi \epsilon_0 \epsilon_r r)$ where
$Z_i$ and $Z_j$ are respectively the valences of particle $i$ and $j$, 
$\epsilon_0$ is the permittivity in vacuum,
$\epsilon_r$ is the dielectric constant of the medium.
It can be re-expressed as 
\begin{equation}
U_{coul}(r)=\frac{Z_i Z_j \lambda_B k_B T}{r}
\label{eqcoulomb2}
\end{equation}
where $k_{B}$ is the Boltzmann constant, $T$ is the temperature, and
$\lambda_B=e^2/(4\pi \epsilon_0 \epsilon_r k_B T)$ is the Bjerrum length,
which denotes the distance between two unit charges at which the Coulomb energy is equal to 
the thermal energy $k_B T$.  
The third interaction is the bond connectivity interaction.
Two adjacent monomers are connected by a \textit{spring},
modeled by a finitely extensible nonlinear elastic potential
\begin{equation}
U_{bond}(b)= -\frac{1}{2} k_{b} b_{max}^2 \ln \left(1-\frac{b^2}{b_{max}^2}\right)
\end{equation}
where $b$ is the length of a bond,
$b_{max}$ is the maximum bond extention, and
$k_{b}$ is the spring constant.
The forth interaction is a harmonic angle potential, which gives 
chain stiffness, 
\begin{equation}
U_{angle}(\theta)= k_{a}( \theta - \theta_0 )^2
\end{equation}
where $\theta$ is the angle between two consecutive bonds, 
$\theta_0$ is the equilibrium angle,
and $k_{a}$ is the force constant.

We perform Langevin dynamics  simulations in this study.
The equation of motion of a particle $i$ is described by the stochastic 
differential equation
\begin{equation}
 m_i \ddot{\vec{r}}_i = -\frac{\partial U}{\partial \vec{r}_i} -m_i\gamma_i \dot{\vec{r}}_i +\vec{{\eta}}_i(t)
\label{eqlangevin}
\end{equation}
where $m_i$ is the mass of particle $i$, $m_i\gamma_i$ denotes the friction coefficient 
and $\vec{\eta}_i$ is a random force modeling the collisions by solvent molecules.
$\vec{\eta}_i(t)$ has zero mean over time and satisfies 
the fluctuation-dissipation theorem: 
\begin{equation}
\left< \vec{\eta}_i(t) \cdot \vec{\eta}_j(t') \right> = 6 k_B T m_i\gamma_i\delta_{ij} \delta(t-t')
\end{equation}
where $\delta_{ij}$ and $\delta(t-t')$ are the Kronecker delta and the Dirac delta function, respectively.
An advantage of using Langevin dynamics is that the temperature control is naturally incorporated. 

In this work, we assume that all the particles have identical mass $m$,  diameter $\sigma$, and 
LJ interaction strength $\varepsilon_{LJ}$.
We set $\lambda_B=3\sigma$ and control temperature at $k_B T=1.2\varepsilon_{LJ}$. 
Coulomb interaction is computed using the technique of Ewald sum.
$k_{b}$ is chosen to be $5.8333 k_B T/\sigma^2$ and $b_{max}$ to be $2\sigma$. 
Under this setup, the average bond length is $1.1\sigma$~\cite{stevens95,hsiao06a}.
We study the effect of chain stiffness by altering the spring constant $k_{a}$ from $0$ to $100 k_B T/rad^2$.
The equilibrium angle $\theta_0$ is set to be $\pi$.
The size of the simulation box is $53.133\sigma$ and the concentration of tetravalent salt ranges 
from $0.0$ to $0.01024 \sigma^{-3}$.
All the particles are subjected to a friction force proportional to the particle velocity. 
We set the damping constant $\gamma$ to $15.0 \tau^{-1}$ to mimic roughly 
an aqueous environment where $\tau= \sigma \sqrt{m/(k_BT)}$ is the time unit.
Simulation time step $\delta t$ is chosen equal to $0.005 \tau$.
The initial chain configuration begins with an extended structure.
An equilibrium phase takes $3 \times 10^6$ to $5 \times 10^7$ time steps
and a production run $10^7$ to $3\times10^8$ time steps, 
depending on chain stiffness and salt concentration.
We found that in the region of intermediate salt concentration,
the transition between different chain structures is slow.
In order to sample more completely different structures, 
five to ten independent runs are performed starting with 
different initial configurations.
The simulations were run using LAMMPS 
package~\footnote{Refer to web site http://lammps.sandia.gov/ for more information about LAMMPS.}.
For the case of the highest salt concentration, the system contains 7776 charged particles.

\section{Results and discussions}
\label{Sec:Result_Discussion}
\subsection{Radius of gyration}
\label{Sec:Rg}
\begin{figure}
\begin{center}
\includegraphics[height=\textwidth,angle=90]{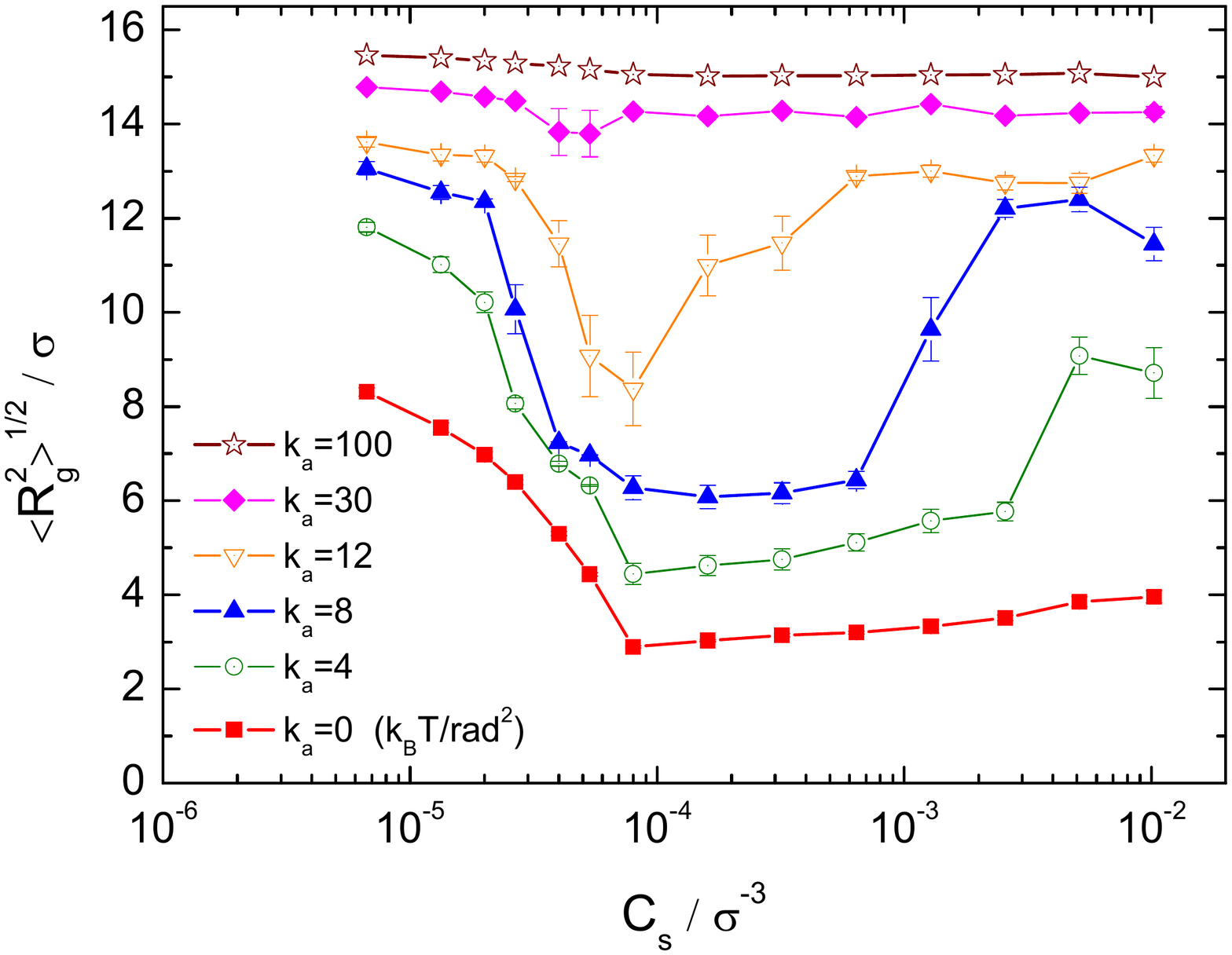}
\caption{}
\label{fig:Rg_line_vCs}
\end{center}
\end{figure}

The size of a polyelectrolyte can be characterized by the radius of gyration, $R_g$. 
It is calculated by taking the square root of the formula, 
\begin{equation}
R_g^2= \frac1N \sum_{i=1}^N (\vec{r}_i-\vec{r}_{cm})^2
\end{equation}
where $N$ is the number of monomers on the chain,
$\vec{r}_{i}$ is the position vector of monomer $i$,
and $\vec{r}_{cm}$ is the center of mass of the chain.
We studied the effects of salt concentration and 
chain stiffness on the root-mean-square radius of gyration,
$\left<R_g^2\right>^{1/2}$. 
The results are shown in Fig.\ref{fig:Rg_line_vCs}.

We observed that the variation of $\left<R_g^2\right>^{1/2}$ behaves
in three distinct manners. 
The first behavior applies to the chain with small $k_{a}$,
such as $k_{a}=0.0 k_B T/rad^2$. 
$\left<R_g^2\right>^{1/2}$ firstly decreases with increasing 
$C_{s}$, the concentration of the tetravalent salt, until the total charge 
of the tetravalent cations is equal to the negative charge  
carried on the chain backbone.  
We  call this particular salt concentration \textit{the equivalence point}.
In this work, the equivalence point occurs at 
$C_{s}= C_{s}^*\equiv 8\times 10^{-5}\sigma^{-3}$ 
because the monomer concentration is fixed at $C_{m}=3.2 \times 10^{-4}\sigma^{-3}$.
$\left<R_g^2\right>^{1/2}$ then increases slowly 
while $C_{s}$ surpasses the equivalence point. 
Therefore, the chain undergoes two structural transitions upon addition of 
tetravalent salt: a shrinking transition and a followed swelling transition. 
The increasing rate of $\left<R_g^2\right>^{1/2}$ in the chain swelling region
is apparently lower than the decreasing rate in the chain shrinking region. 
As a result,  the curve of $\left<R_g^2\right>^{1/2}$ looks similar to a tilted `L' in the semilog plot.
Notice that at high salt concentration, the chain is not swollen back to its original size 
in the absence of salt.  These phenomena have been recently studied in detail by molecular dynamics 
simulations~\cite{hsiao06a, hsiao06b}. 

The second behavior applies to the chain with intermediate $k_{a}$, for
instance, $k_{a}= 8.0 k_B T/rad^2$.
At the beginning of addition of salt, the chain size decreases gently. 
The gentle decrease turns to become a sharp decrease once $C_{s}$ is increased 
beyond some critical value $C_c$, and quickly, $\left<R_g^2\right>^{1/2}$ reaches 
a small constant value.
If $C_{s}$ is further increased over a second critical value $C_d$, 
an abrupt increase in $\left<R_g^2\right>^{1/2}$ takes place.
The equivalence point $C_{s}^*$ is situated in the region between $C_{c}$ and $C_{d}$.  
In the high-salt region $C_{s}>C_{d}$, the chain reattains roughly its original size 
before the condensation occurred.
These two sudden changes suggest that the chain morphology is altered, 
from an extended structure to a condensed structure, and vice versa, 
in a discontinuous way. 
The chain possibly commits a first-order conformational 
transition~\cite{yoshikawa96,noguchi98}.
In the semilog plot, the shape of the $\left<R_g^2\right>^{1/2}$ curve looks like 
the capital letter `U'.
A relevant phenomenon has been observed in the experiments of DNA precipitation by 
polyamines where DNA percentage presented in the supernatant shows a similar 
U-shaped variation as a function of spermidine concentration \cite{pelta96a,raspaud98}.
Suppose that the occurrence of the chain precipitation is directly related to the chain 
size in the solution. 
This variation will then reflect how the dimension of a semiflexible DNA 
varies with salt concentration. 
Our results are in agreement with the experimental observation.

The third behavior is observed for the chain with large $k_{a}$, for example, 
$k_{a}=100 k_B T/rad^2$. 
The influence of the adding salts upon the chain size is not significant.
$\left<R_g^2\right>^{1/2}$  stays essentially at a constant value, 
and in the figure, the curve is roughly a straight line. 

Based upon the variation of chain size against $C_{s}$,
a polyelectrolyte of fixed chain length can be classified into three categories 
according to the chain stiffness:
\begin{itemize}
\item[(1)] \textit{flexible chain}, characterized by a tilted-L-shaped $\left<R_g^2\right>^{1/2}$ curve;
\item[(2)] \textit{semiflexible chain}, characterized by a U-shaped  $\left<R_g^2\right>^{1/2}$ curve;
\item[(3)] \textit{rigid chain}, characterized by an almost straight $\left<R_g^2\right>^{1/2}$ curve.
\end{itemize}
One way to justify if a chain belongs to the category \textit{rigid chain} or not is
to compare the bare persistence length with the chain length.
If the bare persistence length is larger than the chain length, 
the chain is considered to be \textit{rigid}.
In our study, the bare persistence length $\ell_{p,0}$ is related to the 
force constant $k_{a}$ of the harmonic angle potential by 
$\ell_{p,0}=2 k_{a}\langle b \rangle /(k_{B}T)$ where $\langle b \rangle$ is the average
bond length.
Therefore, $\ell_{p,0}$ becomes larger than the chain length while $k_{a}> 24 k_{B}T/rad^2$.
This simple criterion works reasonably well because the chain shows a rigid rod structure 
at all $C_{s}$ for $k_{a}=30 k_B T/rad^2$ and $100 k_B T/rad^2$ 
but not for the cases with smaller $k_{a}$, as having been shown in Fig.\ref{fig:Rg_line_vCs}. 

In the simulations, we observed that, no matter how stiff the chain is,
the polyelectrolyte is in its most stretched state while no salt is added 
into the solution.  
We found that at a given salt concentration, a stiffer chain 
displays a larger $\left<R_g^2\right>^{1/2}$, and hence, occupies 
a larger space.
The results obtained here also demonstrate that the chain stiffness plays an imperative 
role in the determination of the type of the coil-globule transition to be continuous or 
discrete, both in the condensation and in the reexpansion of a polyelectrolyte.

\subsection{Degree of swelling} 
\label{Sec:cRg}
\begin{figure}
\begin{center}
\includegraphics[height=\textwidth,angle=90]{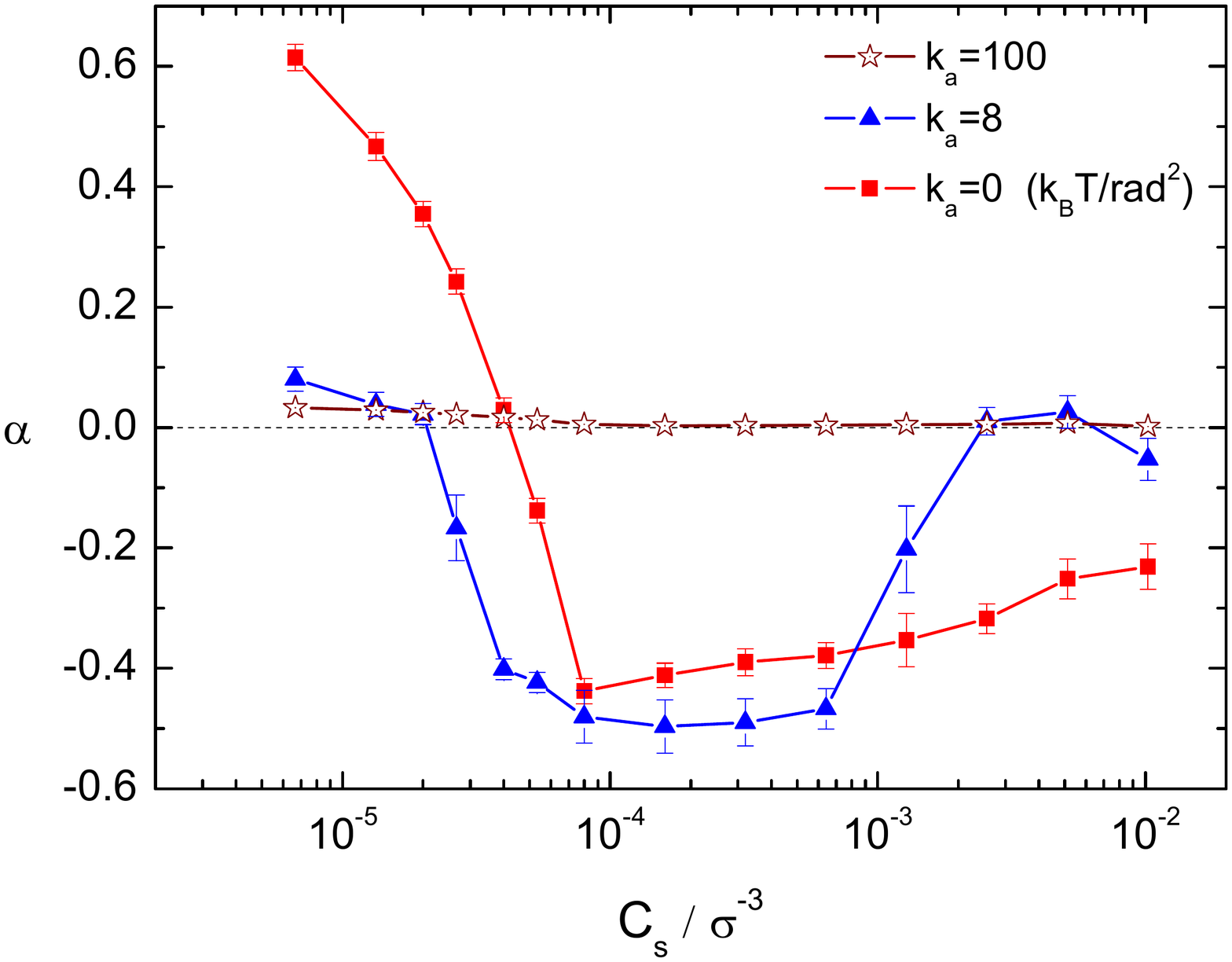}
\caption{}
\label{fig:beta_line_vCs}
\end{center}
\end{figure}

In order to understand the degree of swelling of a polyelectrolyte in salt solutions, 
we calculate the size of the uncharged counterpart of the polyelectrolyte in a salt-free
solution and use it as the reference of comparison. 
The root-mean-square radius of gyration of the uncharged chain, $\left<R_{g,n}^2\right>^{1/2}$, 
is equal to $5.1(1)$, $12.1(2)$, and $14.97(6)$ for the cases 
$k_{a}=0.0$, $8.0$ and $100.0k_B T/rad^2$, respectively. 
We define the degree of chain swelling  by 
$\alpha=(\left<R_g^2\right>^{1/2}/\left<R_{g,n}^2\right>^{1/2})-1$ 
and present the results in Fig.~\ref{fig:beta_line_vCs}.

We observed that in the low-salt region, $\alpha$ is positive. 
It means that  the charged chain occupies a larger space than the uncharged chain,
apparently resulting from the electrostatic repulsion between monomers on the chain.  
The value of $\alpha$ for the flexible chain ($k_a=0.0k_BT/rad^2$) is much larger than the ones 
for the semiflexible chain ($k_a=8.0k_BT/rad^2$) and for the rigid chain ($k_a=100.0k_BT/rad^2$). 
Therefore, the major contribution to the dimension of a semiflexible or a rigid 
polyelectrolyte comes from the intrinsic chain stiffness, whereas the size of a 
flexible chain is mainly determined by the Coulomb repulsion on the chain backbone.  
In the mid-salt region around the equivalence point, 
the degree of swelling is negative for the flexible chain and for 
the semiflexible chain.
The chains are, hence, in collapsed states.
In this region,  electrostatics dominates the chain stiffness.   
The maximum shrinkage, in reference of the size of the neutral chain, can be as large as $50\%$.
For the rigid chain, the collapsing transition does not happen,
and thus, $\alpha$ is close to zero. 
In the high-salt region, we saw that the value of $\alpha$ approaches to zero
for the semiflexible chain. 
The polyelectrolyte hence occupies a space volume similar to its uncharged counterpart.  
For the flexible chain, $\alpha$ is increased with $C_{s}$ but still takes a finite negative value. 
In our previous study~\cite{hsiao06b}, flexible polyelectrolytes in the region of much higher 
salt concentration have been investigated, and the results showed that the chain attains 
a size similar to the neutral polymers.  
Therefore, $\alpha$ is expected to tend toward zero if $C_{s}$ was increased to a much higher 
value in the current study. 

\begin{figure}
\begin{center}
\includegraphics[height=\textwidth,angle=90]{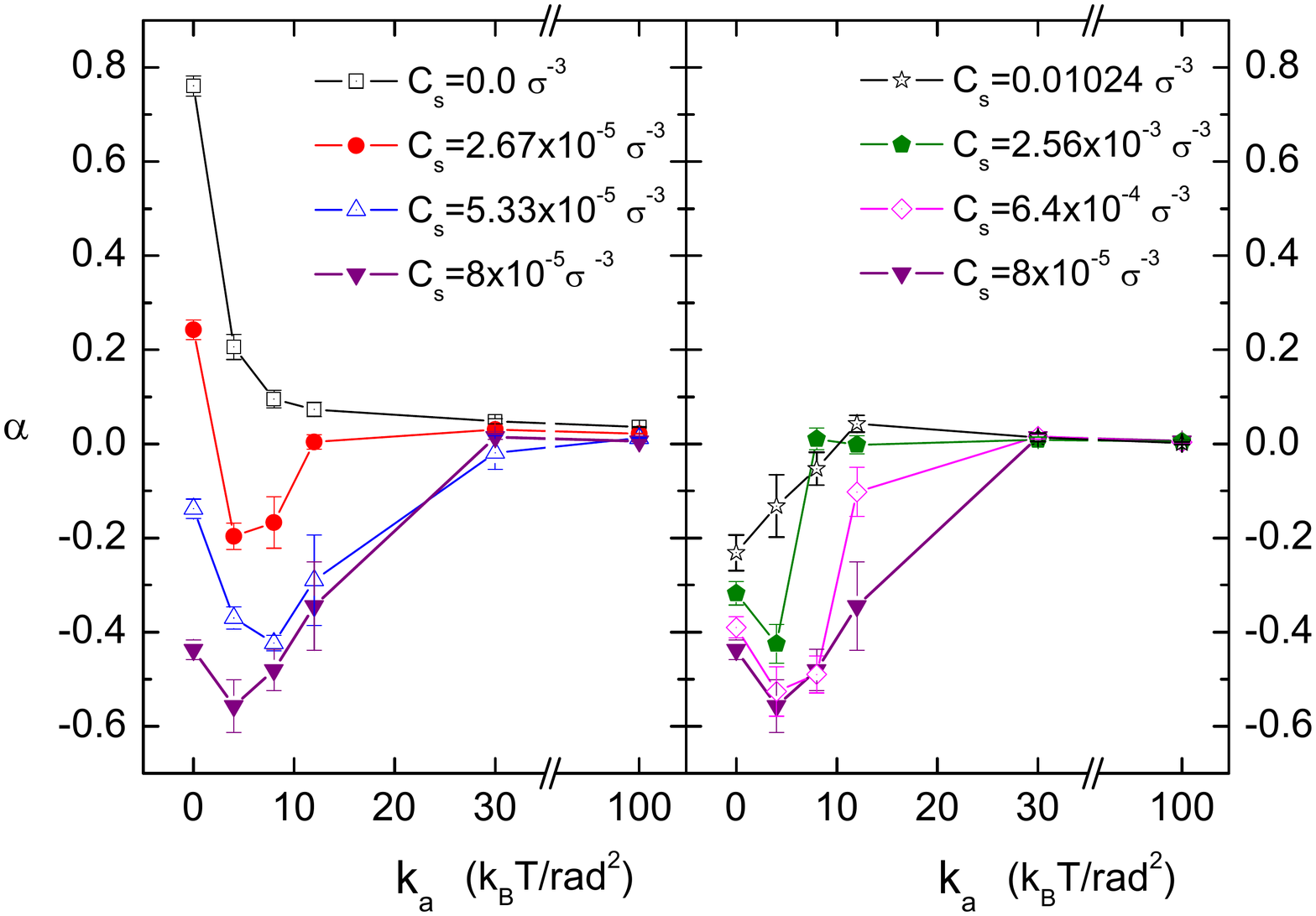}
\caption{}
\label{fig:beta_line_vka}
\end{center}
\end{figure}

In Fig.~\ref{fig:beta_line_vka}, we present the variation of $\alpha$ as a function of 
$k_{a}$ at different $C_{s}$.  
At small $C_{s}$, $\alpha$ is positive and monotonically decreases to zero 
with increasing  $k_a$. 
Opposite trend of behavior is observed at large $C_{s}$, for instance,
$C_{s}=0.01024\sigma^{-3}$,  where $\alpha$ is negative and monotonically increases 
to zero.
In the mid-salt region, $\alpha$ displays a V-shaped curve and the maximum shrinkage is happened 
around $k_{a}=5k_B T/rad^2$.    
Thus, by tuning $k_a$ and varying $C_{s}$, one is able to control 
the degree of chain swelling to design desired functions in real applications. 

The results obtained here tell us that in a salt-free solution, 
the larger the value of $k_a$, the less significant the size difference 
between a charged chain and its uncharged counterpart will be.
Since at large $C_{s}$ a chain occupies a space similar to the one of a 
neutral polymer, a semiflexible chain will reattain roughly its original size
at zero $C_{s}$  by addition of many salts into the solution,
but a flexible chain will not reattain it.

\subsection{End-to-end distance} 
\label{Sec:Re}
\begin{figure}
\begin{center}
\includegraphics[height=\textwidth,angle=90]{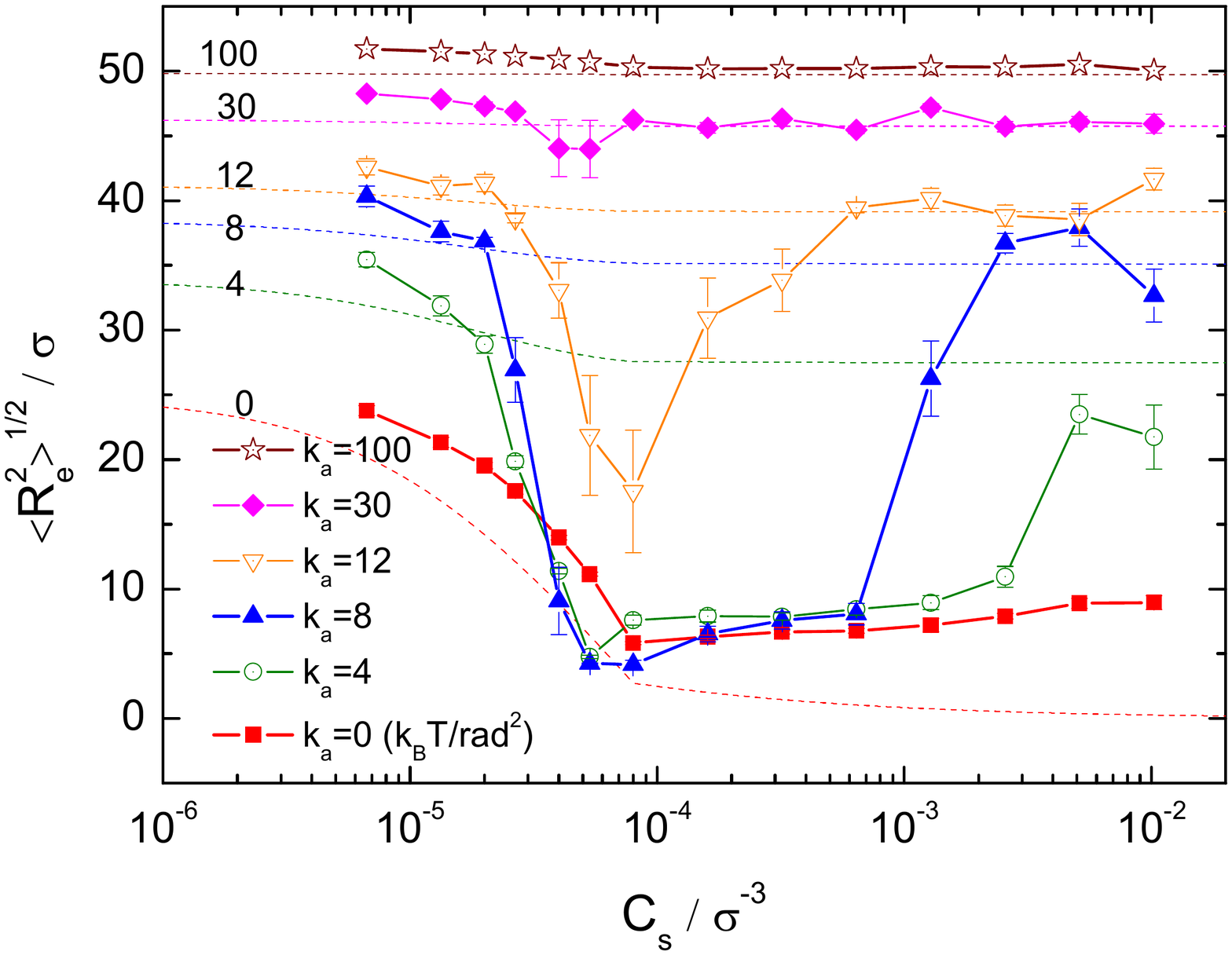}
\caption{}
\label{fig:Re_line_vCs}
\end{center}
\end{figure}

Conventionally, the end-to-end distance $R_e$ is used to characterize the size of a polymer. 
In Fig.~\ref{fig:Re_line_vCs} we present how the root-mean-square end-to-end distance,
$\left< R_e^2\right>^{1/2}=\left< (\vec{r}_{N}-\vec{r}_1)^2 \right>^{1/2} $, varies with
salt concentration for different chain stiffness, where 
$\vec{r}_{1}$ and $\vec{r}_{N}$ are the position vectors of the two ends of a chain.
The behavior of $R_e$ resembles the behavior of $R_g$:
a flexible chain displays a tilted L-shaped $R_e$ curve,
a semiflexible one displays a U-shaped $R_e$ curve,
and a rigid one displays a straight $R_e$ curve.
Nonetheless, the value of $R_e$ does not precisely reflect the 
\textit{real} size of a chain. 
For example,  $R_e$ for a semiflexible chain becomes smaller than
that for a flexible chain in the mid-salt region,   
although a semiflexible chain has a larger size (cf.~Fig.~\ref{fig:Rg_line_vCs}).
This inconsistency results from the formation of ordered structures
which gives a small $R_e$, such as hairpin or ring-like structure. 
Therefore, the quantity $R_e$ is not always suitable to characterize the dimension of
a chain in a salt solution, particularly when the chain stiffness is intermediate.

The mean-square end-to-end distance of a chain 
has been derived theoretically under the framework of the worm-like chain (WLC) model 
and reads as
\begin{equation}
\left< R_e^2\right>=2\ell_{p} L - 2 \ell_{p}^2 \left(1-\exp(-L/\ell_p) \right)
\label{formula:Re2}
\end{equation}
where $L$ is the contour length and $\ell_p$ is the persistence length~\cite{rubinstein03book}.
For a polyelectrolyte chain, $\ell_p$ is the sum of two contributions: 
the first one is the bare persistence length $\ell_{p,0}$, coming from the intrinsic chain 
stiffness, and the second one is the electrostatic persistence length $\ell_e$, 
due from the electrostatic interaction. 
Odijk~\cite{odijk77} and, independently, Skolnick and Fixman~\cite{skolnick77} (OSF) 
have deduced a formula for the electrostatic persistence length 
in the limit of stiff chain and low ionic strength. 
This formula reads as 
\begin{equation} 
\ell_{e}^{OSF}=\frac{\xi^2}{4\kappa^2 \lambda_B}
\label{formula:le}
\end{equation}
where $\xi$ is the Manning parameter, defined as the ratio of $\lambda_B$
over the effective charge distance on the chain, and $\kappa$ is the inverse Debye length. 
In order to make a comparison of the simulation results with the theory, we adopted the 
Manning-Oosawa (MO) condensation theory~\cite{manning69,oosawa71} to estimate $\xi$.
In our case, $\xi$ is renormalized to a value $1/4$, 
due to the condensation of tetravalent counterions,
while $C_{s} \ge C_{s}^* \equiv C_{m}/4$.
For small salt concentration $C_{s}<C_{s}^*$, all the tetravalent counterions 
are expected to condense on the chain, but the number of these ions is not sufficient 
to renormalize $\xi$ to $1/4$. 
At this moment, $\xi$ decreases linearly from 1 to $1/4$ with $C_{s}$,
or equivalently, $\xi =1 - 3C_s/C_m$. 
Plugging $\xi$ and $\kappa^2=4\pi\lambda_B (C_m + 4C_s + 4^2C_s)$ into Eq.~(\ref{formula:le}),
we obtained $\ell_e^{OSF}$, and then, added it to $\ell_{p,0}=2 k_{a}\langle b \rangle /(k_{B}T)$ 
to get the persistence length $\ell_p$.  
Finally, using Eq.~(\ref{formula:Re2}), we calculated $\left<R_e^2\right>^{1/2}$
and the results for different chain stiffness are plotted in dashed curves in 
Fig.~\ref{fig:Re_line_vCs}. 
We observed that the WLC model, combining with the OSF and the MO theories, gives a 
qualitative prediction for $\left<R_e^2\right>^{1/2}$ at low salt concentration 
for semiflexible and stiff chains.
Surprisingly, it also gives a not-bad description of $\left<R_e^2\right>^{1/2}$
for the flexible chain with zero $k_{a}$ and even captures the decreasing behavior 
 up to $C_{s}=C_{s}^*$, although the OSF 
and the MO theories are not expected to work well neither for flexible chains 
nor for a solution with  multivalent salt.
The theoretical value is found to be smaller than the value obtained from the simulations.
This deviation can be attributed to the effect of the excluded volume 
interaction, which is neglected in the derivation of the WLC model
and should give a positive contribution to the chain size if it is taken
into account.  In the mid-salt regime, the theory fails to predict 
$\left<R_e^2\right>^{1/2}$ for semiflexible chains.
In this region, the chain is condensed by tetravalent counterions.
The small chain size suggests that the persistence length is 
significantly reduced, and as a consequence, the electrostatic persistence 
length takes a negative value. 
The negative electrostatic persistence length has been experimentally 
observed~\cite{baumann97}.
At high salt concentration, the WLC model again describes well the 
end-to-end distance except for the case with zero $k_a$; 
at this moment, $\ell_e^{OSF}$ is almost zero and the persistence length is
dominated by the bare persistence length.

\subsection{Snapshots of simulation}
\label{Sec:snapshot}
Before processing a more quantitative analysis, 
we present in this section snapshots of our simulations
to give readers a vital illustration how the chain looks like
in different regions of salt concentration and chain stiffness.

\begin{figure}
\begin{center}
\includegraphics[width=0.5\textwidth]{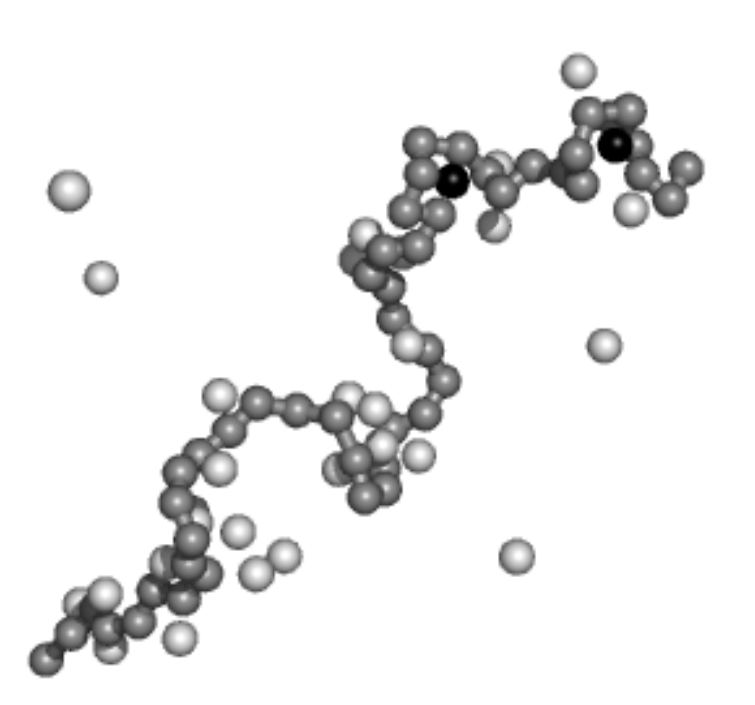}
\\(a)\\
\vspace{1cm}
\includegraphics[width=0.5\textwidth]{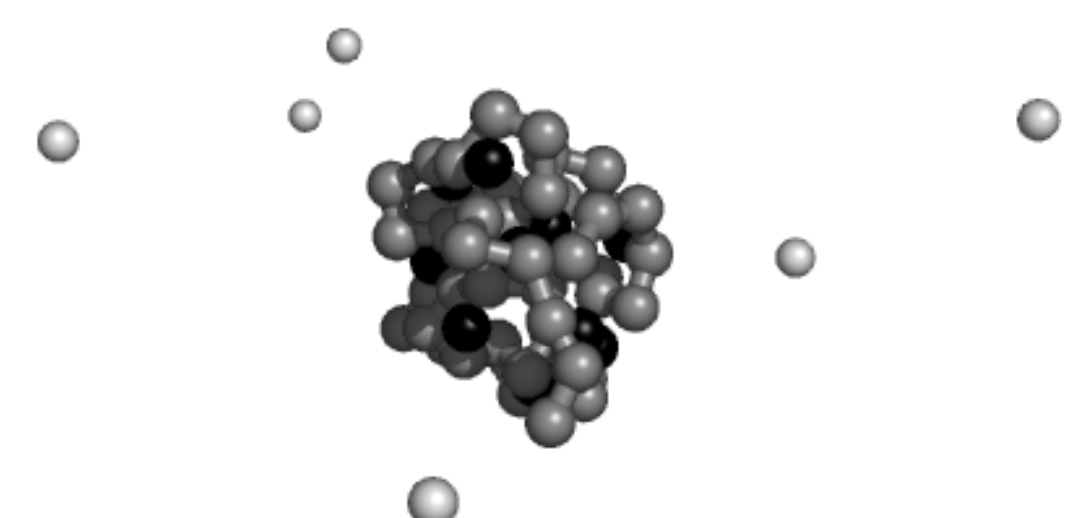}
\\(b)\\
\vspace{1cm}
\includegraphics[width=0.5\textwidth]{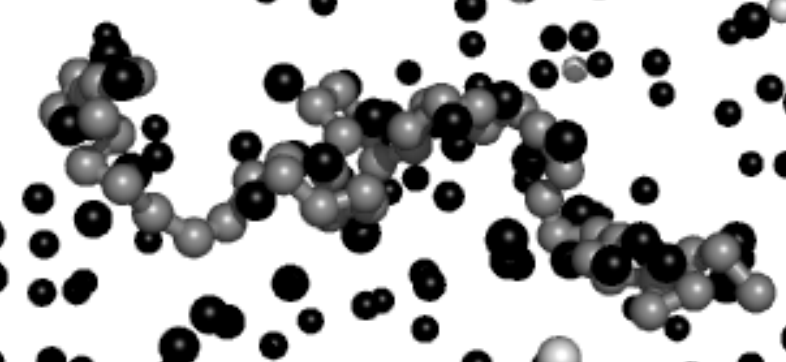}
\\(c)\\
\vspace{1cm}
\caption{}
\label{fig:flexiblechain}
\end{center}
\end{figure}

Fig.~\ref{fig:flexiblechain} shows snapshots of a flexible 
chain at low, middle, and high salt concentrations.
The morphology of the chain is elongated at low salt concentration,
and displays large-angle turns near the places where the counterions 
condense, as shown in Fig.~\ref{fig:flexiblechain}(a). 
In the mid-salt region around the equivalence point,
the chain collapses and exhibits a randomly-arranged compact structure 
(see Fig.~\ref{fig:flexiblechain}(b)). 
At high salt concentration (Fig.~\ref{fig:flexiblechain}(c)),
the chain swells and acquires a zigzagged, extended structure, 
but not as elongated as  at low salt concentration. 

\begin{figure}
\begin{center}
\includegraphics[width=0.5\textwidth]{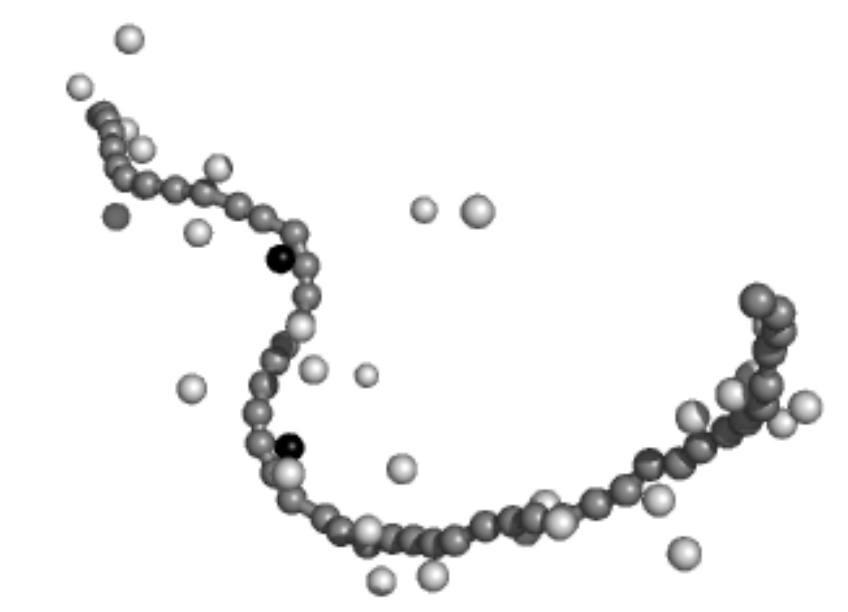}
\\(a)\\
\vspace{1cm}
\includegraphics[height=\textwidth,angle=270]{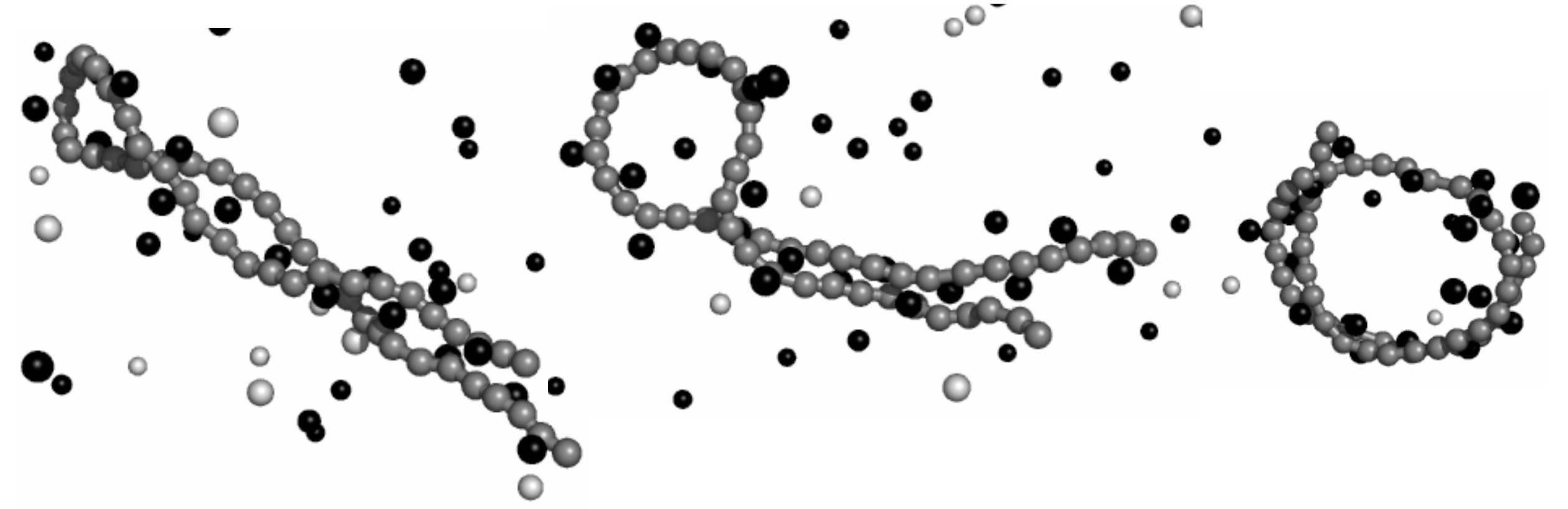}
\\(b)\\
\vspace{1cm}
\includegraphics[width=0.5\textwidth]{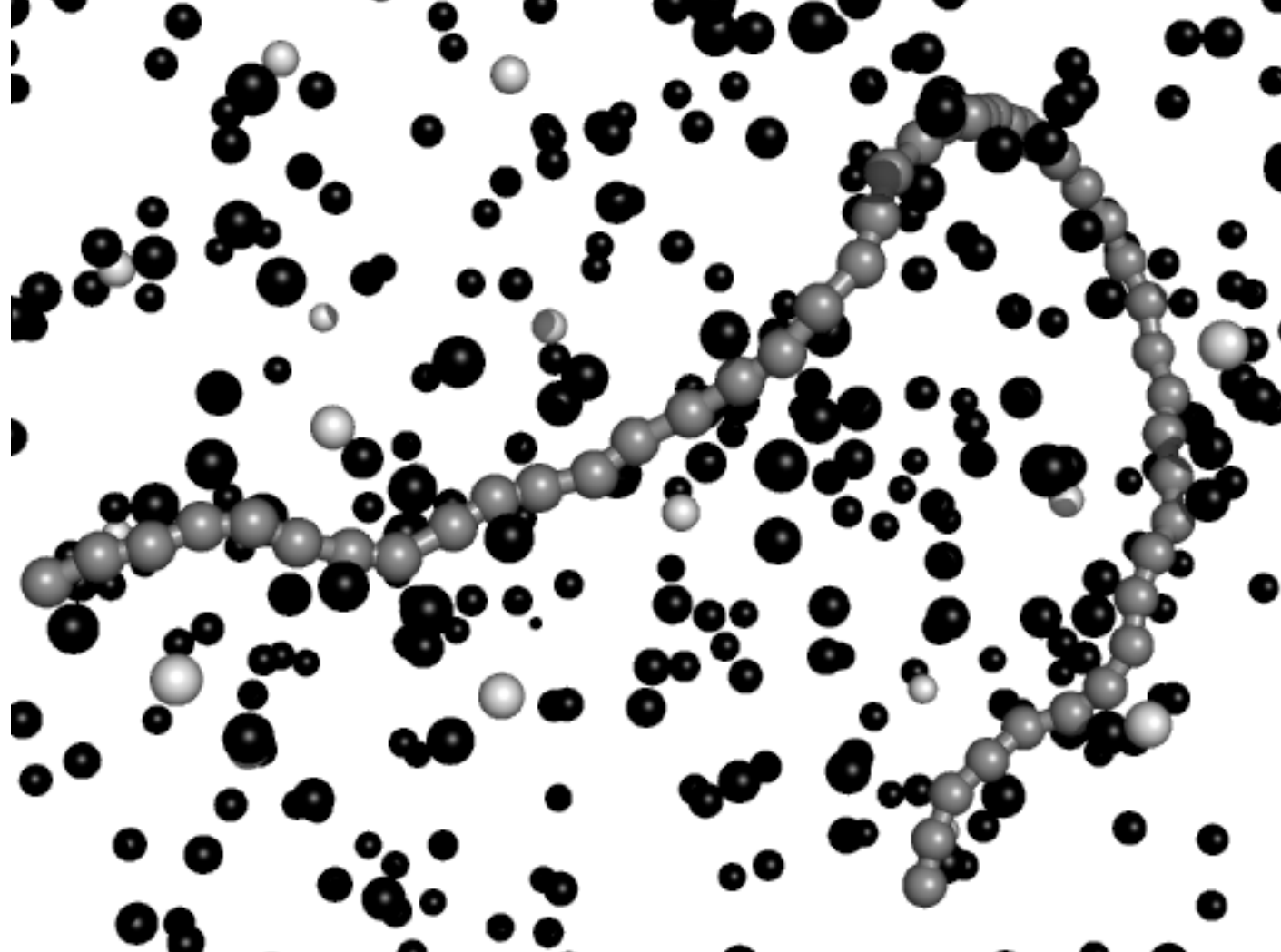}
\\(c)\\
\vspace{1cm}
\caption{}
\label{fig:semiflexiblechain}
\end{center}
\end{figure}

Fig.~\ref{fig:semiflexiblechain} shows snapshots of a semiflexible 
chain at three salt concentrations.
The shape of the chain is smoother than that of the flexible chain.
At low salt concentration (Fig.~\ref{fig:semiflexiblechain}(a)),
the chain is elongated and not zigzagged.
At middle salt concentration, the chain is
condensed, and noticeably, forms ordered structure, in contrast to the 
disordered globule formed by a flexible chain.
Typical ordered structures are shown in 
Fig.~\ref{fig:semiflexiblechain}(b). 
They are, from left to right, hairpin, racquet, and toroid. 
These typical structures have been observed in the experiments of 
DNA condensation~\cite{chattoraj78,martin00,yoshikawa02,maurstad03}.
Unlike the case for flexible chains, a condensed counterion 
can hardly induce a large-angle turn on the chain backbone. 
If it is the case, the turn happens to be the folded end of a 
hairpin structure and counterions are condensed between the two branches
of the hairpin.
The whole structure looks similar to peas in a pod.  
At high salt concentration, the chain displays again an elongated 
structure (see Fig.~\ref{fig:semiflexiblechain}(c)).

\begin{figure}
\begin{center}
\includegraphics[height=0.5\textwidth]{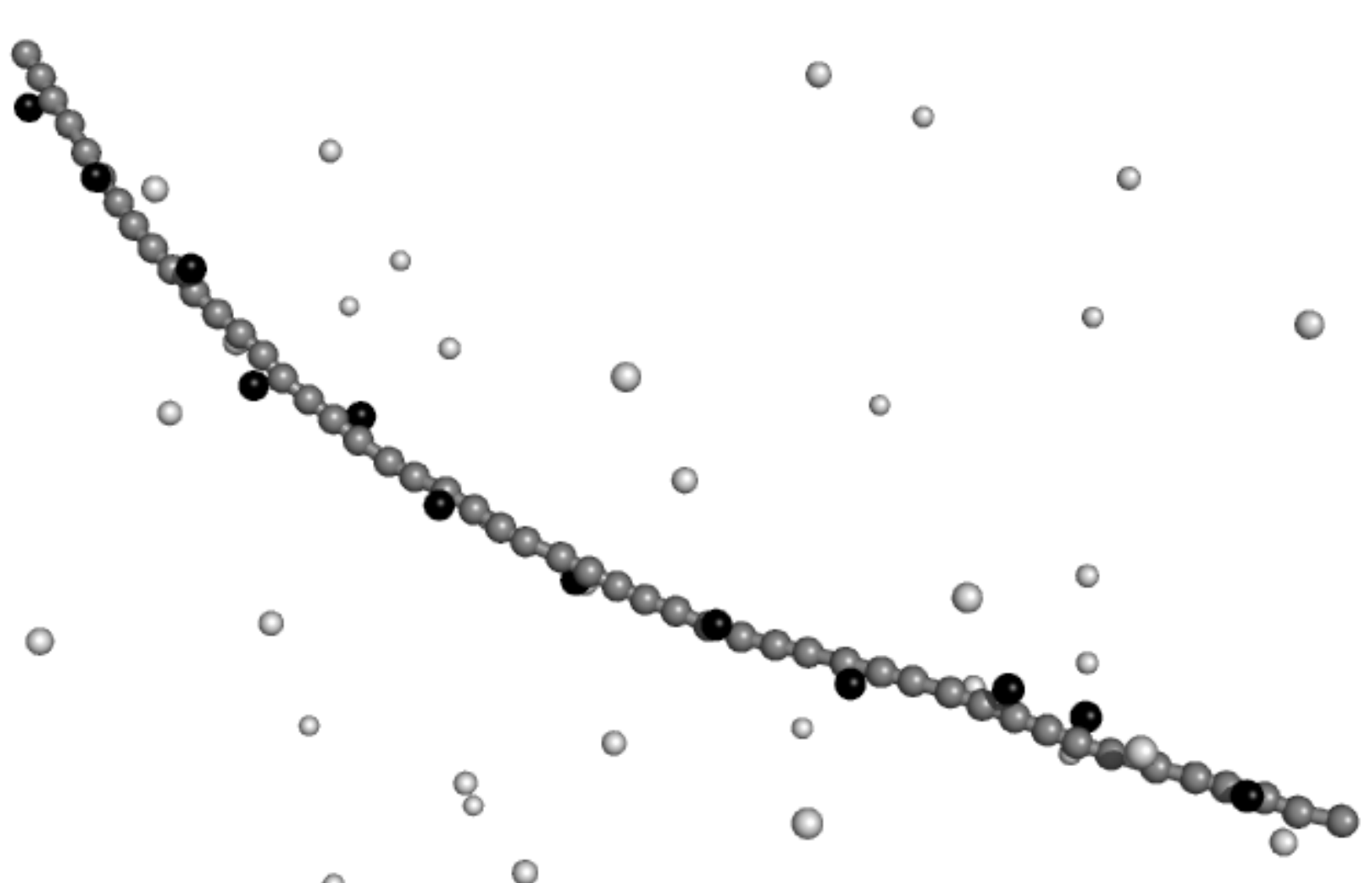}
\caption{}
\label{fig:stiffchain}
\end{center}
\end{figure}

Fig.~\ref{fig:stiffchain} shows the snapshot of a rigid chain.
The chain exhibits a stretched structure no matter how many salts 
are added into the solution.
The chain is stiff enough to resist from bending or twisting. 

From these snapshots, we found that a compact chain structure 
is formed when condensed tetravalent counterions can overcome 
the chain stiffness to bridge between monomers and have enough number 
to tightly bind the whole chain structure. 
We observed that, regardless of the chain stiffness, 
the number of tetravalent counterions condensed on a chain increases
with $C_{s}$. 
When $C_{s}$ is smaller than the equivalence point $C_{s}^*$, 
almost all the tetravalent counterions are condensed on the chain. 
Only when $C_s>C_s^*$, non-condensed tetravalent counterions can appear
in the bulk solution. 
Moreover, the number of condensed tetravalent counterions surpasses 
the number needed to neutralize the bare chain charge when $C_s>C_s^*$,
and as a consequence, the chain is \textit{locally 
overcharged}~\footnote{``Locally overcharged'' does not necessarily 
mean a reversal of the
sign of the effective chain charge because counterions and coions
can form layering organization around a polyelectrolyte at high
salt concentration, resulting in an oscillatory charge 
distribution around a chain. See Refs~\cite{hsiao06a, hsiao06b} 
for detailed discussions for flexible chains.}.
For flexible chains, there are mainly two effects to drive  
chain reentrance at high salt concentration. 
The first one is the Coulomb repulsion between condensed 
tetravalent counterions. 
Since the chain is surcharged by the condensed tetravalent counterions, 
the net repulsion between these ions manifests its effect 
in the swelling of the chain size.  
The second one is the presence of the non-condensed tetravalent 
counterions in the bulk solution. 
These non-condensed counterions interact with the chain, and can dynamically 
replace previously condensed counterions and become newly condensed counterions. 
During this process, the chain morphology oscillates from a loose 
globule to an expanded structure and vice versa. 
Increasing salt concentration increases the concentration of non-condensed
tetravalent counterions in the bulk, which makes easier the replacement of 
tetravalent counterions to take place.  
The chain hence more frequently stays in 
an expanded state and the averaged chain size increases.   
For semiflexible chains, there is, in addition, a third effect taking part
in the reentrant transition --- the chain stiffness.  
The chain stiffness always gives an effect against the formation of
a compact structure.
Together with the instability induced by the presence of non-condensed 
tetravalent counterions, a chain can dramatically leave a condensed state 
to an elongated structure at high salt concentration.

\subsection{Probability density distribution} 
\label{Sec:pdd}
In the previous sections, the conformation of a polyelectrolyte was studied 
by averaged quantities, such as the root-mean-square $R_g$ and 
the root-mean-square $R_e$.
In fact, averaged quantities cannot reflect precisely the chain shape 
since chain morphology varies with time during simulations.
In order to understand the details of chain morphology,
we investigate in this section the probability density distribution (PDD) of 
conformational quantities.

Two quantities are studied: The first one is the asphericity $A$ and the second 
is the radius of gyration $R_g$.
The asphericity $A$ measures the deformation of chain morphology from a spherical geometry
and is defined by~\cite{rudnick86}
\begin{equation}
A=\frac{(\lambda_1-\lambda_2)^2+(\lambda_2-\lambda_3)^2+(\lambda_3-\lambda_1)^2}{2(\lambda_1+\lambda_2+\lambda_3)^2}.
\end{equation}
$\lambda_1$, $\lambda_2$, and $\lambda_3$ are the three eigenvalues of 
the gyration tensor of the chain calculated by
\begin{equation}
{\cal{T}}_{\alpha\beta}=\frac{1}{N}\sum_{i=1}^{N} (\vec{r}_i-\vec{r}_{cm})_{\alpha}(\vec{r}_i-\vec{r}_{cm})_{\beta}
\end{equation}
where the subfixes $\alpha$ and $\beta$ denote the three Cartesian components $x$, $y$ and $z$.
The value of $A$ ranges between 0 and 1.
It is equal to 0 for a perfect sphere, 0.25 for a perfect ring, and 1 for a straight line.
For a coil chain, $\left<A\right>$ is 0.431 obtained from simulation~\cite{bishop88}.

\begin{figure}
\begin{center}
\includegraphics[height=\textwidth,angle=90]{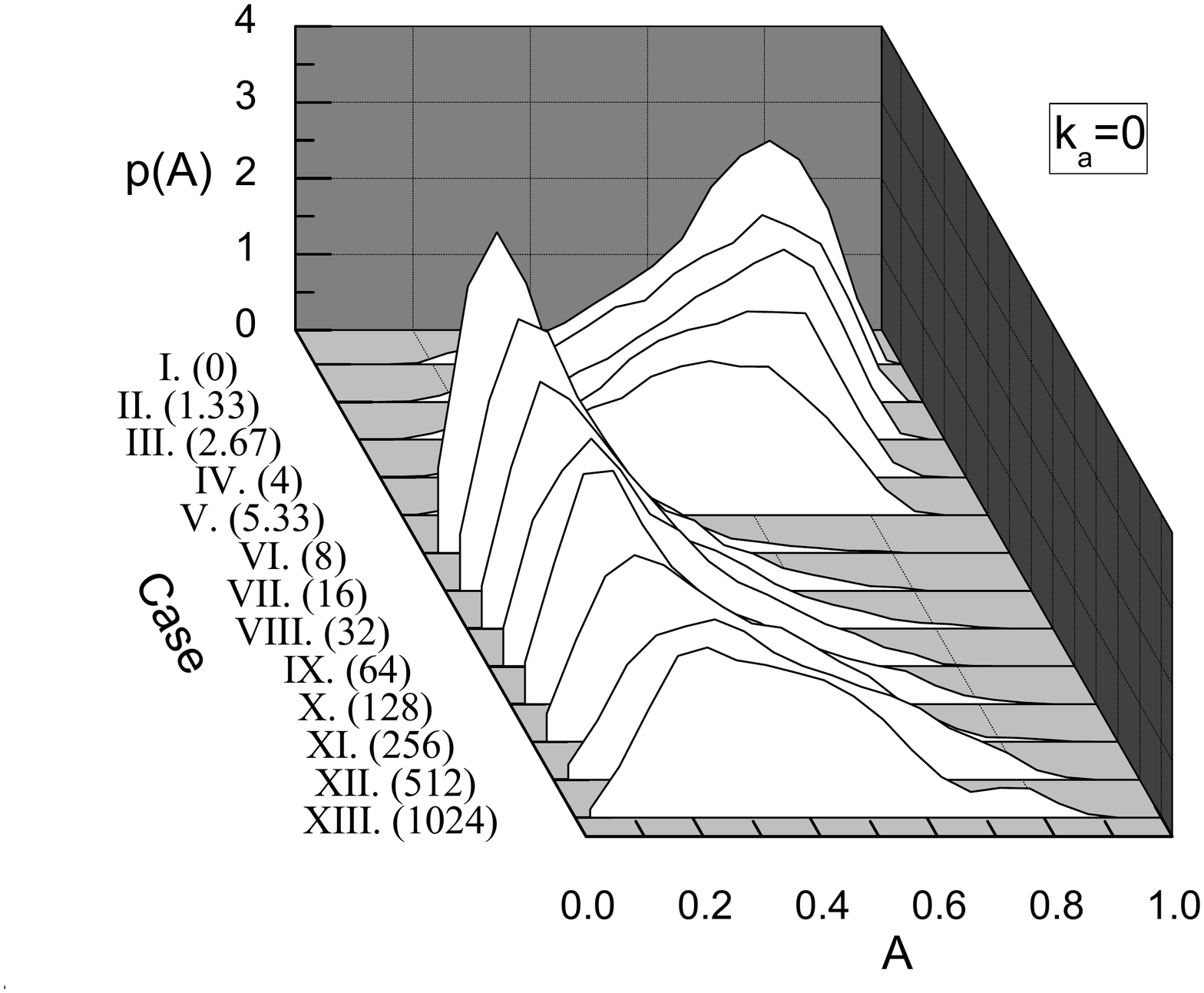}
\caption{(a)}
\label{fig:histoD_ka0}
\end{center}
\end{figure}

In Fig.~\ref{fig:histoD_ka0}, parts (a) and (b), we present PDD of  asphericity, $p(A)$, 
and PDD of  radius of gyration, $p(R_g)$, for the flexible chain with null $k_{a}$.
We observed that $p(A)$ shows a peak near $A=0.8$ in the salt-free solution,
indicating that the chain favors an extended structure.
The peak gradually moves toward a small $A$ as $C_s$ is increased, and
the width of the peak becomes broader.  The width then decreases while $C_s$ 
approaches to the equivalence point $C_{s}^*$ (Case VI in the figure).
At $C_{s}^*$, $p(A)$ shows the most pronounced peak at $A \simeq 0.1$,
and therefore, the favored  morphology is a compact sphere-like structure.
If $C_s$ is further increased, the peak becomes broaden and the position
of the peak shifts not much toward a larger value. 
In this salt region, the chain behaves like a random coil 
and alternates between a globule and an extended structure.  
For $p(R_g)$, similar behavior was observed.
The peak of $p(R_g)$ moves left to a small $R_g$ and then moves slightly right 
as $C_{s}$ is increased. Also, the most pronounced peak appears
at the turning point $C_{s}=C_{s}^*$.  

Figs.~\ref{fig:histoD_ka8}(a) and (b) shows, respectively, $p(A)$ and $p(R_g)$
for the semiflexible chain with $k_a=8k_B T/rad^2$. 
In the low-salt region, $p(A)$ displays a single peak at $A \simeq 0.9$, and therefore, 
the chain favors to exhibit a rod-like structure.
If the salt concentration is increased to some intermediate value, a second peak appears 
at $A\simeq 0.25$.
It indicates the formation of a ring-like (toroid) structure.
The second peaks first increases in its height with $C_s$ and then decreases.  
It disappears while $C_{s}$  surpasses some critical value, and
the chain reattains a rod-like structure.
More information can be obtained if we study $p(R_g)$ at the same moment. 
For example, at $C_s=2.67 \times 10^{-5} \sigma^{-3}$ (Case III in the figure), 
$p(R_g)$ shows a doubly peaked distribution and the second peak is
roughly situated at the middle position between the first peak and $R_g=0$. 
We know from $p(A)$ that the chain exhibits a rod-like structure 
at this salt concentration.  
The second peak in $p(R_g)$, hence, corresponds to hairpin structure 
because hairpin is a two-folded rod and its radius of gyration 
is equal to half of that of the unfolded rod.
Consequently, the chain morphology alternates between unfolded rod
and hairpin structure.
At higher salt concentration, the first peak in $p(R_g)$ disappears 
and a new peak appears at $R_g \simeq 5 \sigma$. 
This new peak is attributed to toroidal structure because 
it corresponds to the peak of $p(A)$ at $A\simeq 0.25$,
which is the typical value for a ring structure.
Therefore, in this salt region coexist 
two condensed structures, hairpin and toroid.
We noted that probability density for $A$ or $R_g$ is nearly zero  between the two peaks
when $C_{s}$ is not large, which indirectly indicates that the structural transition
between hairpin and toroid rarely happens
during  a run. 
If $C_{s}$ is large, for instance, $C_s=6.4 \times 10^{-4} \sigma^{-3}$ , 
a finite value appears between the two peaks,
indicating that the two structures can frequently transit between each other. 
We mention that these results were obtained by doing statistics combined from
several independent runs starting with different initial configurations 
to circumvent the problem of slow transition.
At $C_{s}=0.00128\sigma^{-3}$ (Case X in the figure), $p(R_g)$ shows three peaks,
indicating a coexistence between the two condensed and the extended-chain 
structures.  
One thing is worth to be noticed:
In this middle salt region, the peak in $p(R_g)$, which corresponds to a toroidal structure, 
shifts toward a large value as $C_{s}$ is increased.
It suggests that the toroidal size grows up with salt concentration. 
This phenomenon is in agreement with the  observation of Conwell 
\textit{et al.}~\cite{conwell03}.
In the experiment, they found that monovalent or divalent salt 
causes an increase in diameter of a condensed DNA toroid.  
Finally, if $C_{s}$ goes even higher, $p(R_g)$ displays only one peak  
at $R_g \simeq 13 \sigma$, and hence, the favored shape is an unfolded chain.

We emphasize that the PDDs for the semiflexible chain shows different
behavior, compared with the ones for the flexible chain. 
In the middle salt region, they are multiply peaked and the peaks  
appear or disappear in a sudden way. Moreover, each peak corresponds 
to an ordered structure such as toroid or hairpin.
On the other hand, the PDDs for the flexible chain display only single peaks
and the peak position moves continuously in the space with salt concentration.
As a consequence, the coil-globule transition for the semiflexible chain occurs 
in a discrete manner, whereas in a continuous way for the flexible chain. 
Both the discrete and the continuous transition have been demonstrated 
in simulations by other authors~\cite{noguchi98,ivanov00,sarraguca03,khan05}.
However, the subjects were focused on the chain collapsing from
a coil state to a globule state, happened at low salt concentration or
at low temperature.    
Our study presented here goes further and gives a thorough description 
of how the PDDs varies with salt concentration, not only in the low
salt region but also in the high salt region. 
The sudden appearance and disappearance of peaks show that the discrete 
nature of transition takes place also at high salt concentration 
when a semiflexible chain redissolves from an ordered globule state 
to a coil state. 
To our knowledge, this is the first time such observation
at high $C_{s}$ is reported. 

We present in Fig.~\ref{fig:histoD_ka100}(a) and Fig.~\ref{fig:histoD_ka100}(b)
$p(A)$ and $p(R_g)$, respectively, for a rigid chain with $k_{a}=100k_B T/rad^2$.
$p(A)$ shows a sharp peak near $A=1$ and $p(R_g)$  a 
sharp peak at $R_g$ around $15\sigma$ for all the salt concentration 
investigated.
The chain exhibits a fully extended structure. 
The results can be verified by estimating the chain length 
from the radius of gyration.
It is known that the length $\ell$ of a rod is related to
the radius of gyration by $R_g^2=\ell^2/12$.  
Consequently, $R_g=15\sigma$ yields $\ell$ equal to $51.96\sigma$, 
which is in good agreement with the length of a fully extended
chain since our chain has 47 bonds and the average
bond length is $1.1\sigma$.

We remark that the results presented here were obtained 
from long simulations.  
We have calculated the correlation time for the quantities $A$ and $R_g$ 
at each salt concentration. 
The results shows that our simulations contains hundreds to 
thousands independent data for the cases of flexible chain and rigid chain.
Therefore, we are confident that Fig.~\ref{fig:histoD_ka0} and 
Fig.~\ref{fig:histoD_ka100} report correct equilibrium populations.  
For the case of semiflexible chain, particularly in the 
mid-salt region, the system shows rich morphologies and the transition between 
different morphologies is slow. We have preformed several independent runs to
sample different trajectories in order to improve the statistics. The analysis 
of the correlation time shows that at least, thirty independent samples were 
cumulated in the simulations.  
The results reported in Fig.~\ref{fig:histoD_ka8}, hence, may not be very 
accurate. 
However, the PDDs show consistent trend of variation against $C_{s}$, and also 
in Fig.~\ref{fig:Rg_line_vCs}, no remarkably large fluctuation is appeared 
in the $\left<R_g^2\right>^{1/2}$ curve. 
Therefore, these results still give an appreciable description of 
the equilibrium populations.

\subsection{Dynamics of conformation} 
\label{Sec:x-t}
We have shown that a polyelectrolyte can be condensed 
by multivalent salt.  Particularly, when the chain is semiflexible, the 
condensed chain displays an ordered structure, such as hairpin or toroid. 
In order to get insight of the phenomena, we take the semiflexible chain of $k_a=8k_BT/rad^2$ 
as an example, and study in this section how the conformational quantities, 
$A$, $R_g$ and $R_e$, evolve with time.

We focus at first the conformational dynamics at $C_{s}=8\times 10^{-5} \sigma^{-3}$.
In Fig.~\ref{fig:histoD_ka8}, we have demonstrated that the PDDs of 
the semiflexible chain are doubly peaked at this salt concentration. 
The chain stays in a hairpin state or a toroid state and the transition 
between the two states is not frequent.
For this case, several independent runs were performed
starting with different initial configurations.
In order to get insight of the dynamics, we show firstly in Fig.~\ref{fig:yt_plot_S8c4} 
the time evolution of $A$, $R_g$ and $R_e$ for one of the independent runs, 
in which the chain exhibits hairpin structure.

The asphericity $A$ fluctuates steadily around $0.9$ as shown in Fig.~\ref{fig:yt_plot_S8c4}(a). 
At the same moment, the radius of gyration fluctuates around $R_g=6.5\sigma$ and 
$R_e$ fluctuates around a small value (see Fig.~\ref{fig:yt_plot_S8c4}(b)). 
Therefore, the chain has its two ends fluctuating near the open end 
of the hairpin structure. 
Readers can see from the figures that this structure is 
lasted for a duration of at least $2\times 10^8$ simulation steps in this run. 
Please be aware  that $2\times 10^8$ simulation steps is a long time period 
in a typical molecular simulation. 
In our case, it is approximately equal to a duration of $2\mu s$, provided that  
$\sigma$ is $2.4${\rm \AA}, 
$T$ is 300{\rm K} and 
$m$ is 200{\rm Da}.

We then show in Fig.~\ref{fig:yt_plot_S8c5} the dynamics of these quantities 
when the chain exhibits toroidal structure.
The asphericity $A$ fluctuates around $0.25$ except at four places,  
fluctuating around $0.7$ (cf.~Fig.~\ref{fig:yt_plot_S8c5}(a)). 
At these four places, the chain displays a twisted ring structure, 
similar to the shape of the digit `8'.
Since the two ends of a toroidal chain is generally not very close to
each other, $R_e$ fluctuates around a large value (see Fig.~\ref{fig:yt_plot_S8c5}(b)), 
compared to the value for hairpin structure (Fig.~\ref{fig:yt_plot_S8c4}(b)). 
Resembling the previous run, the toroid structure is continued for nearly 
$2\times 10^8$ simulation steps.  
The morphological transition between toroid and hairpin is, hence, 
not frequent at this salt concentration. 

The transition between different ordered structures becomes observable in simulations 
if the salt concentration is high. 
For example, at $C_s=6.4 \times 10^{-4} \sigma^{-3}$ (shown in Fig.~\ref{fig:yt_plot_S96}(a)), 
the asphericity fluctuates around two values, at $A=0.25$ and at $A=0.85$, which  indicates 
the transition between ring-like and rod-like structures.
Moreover, Fig.~\ref{fig:yt_plot_S96}(b) shows that 
the radius of gyration associated to the rod-like structure is larger than 
that associated to the ring-like structure, 
but the end-to-end distance is smaller.  
It enables us to identify this rod-like structure to be a two-folded chain 
(\textit{i.e.} hairpin), which was also confirmed from the simulation snapshots. 
The second example is chosen at $C_s=1.28\times 10^{-3} \sigma^{-3}$ 
(see Fig.~\ref{fig:yt_plot_S192}). 
The asphericity shows similar fluctuation around the two values. 
However, the value, $A\simeq 0.85$, which corresponds to rod-like structure, 
is now associated to two behaviors of $R_g$, 
fluctuating at $R_g\simeq 6.5\sigma$ and $R_g\simeq 13\sigma$.
Combining the information obtained from $R_e$, we assert that the first 
fluctuation of $R_g$  corresponds to a two-folded chain structure and 
the second one corresponds to an unfolded-chain structure. 
Therefore, the chain transits between three states, which are unfolded chain state, 
hairpin state, and toroid state.

Recently, dynamic exchange between toroid and folded-chain structures 
in a salt-free solution of semiflexible polyelectrolyte
has been numerically demonstrated by tuning the Coulomb strength 
parameter~\cite{ou05}. 
The work showed a back-and-forth flipping between the two structures, 
similar to our observations.
Our study, moreover, showed that transitions between different
chain states can be enhanced by addition of multivalent salt beyond the 
equivalence point $C_{s}^*$.
The mechanism is understood as follows.
Below $C_{s}^*$, most of the multivalent counterions are 
condensed on the chain.
Roughly while $C_{s}$ is larger than $C_{s}^*$,  multivalent counterions 
start to exist in the bulk solution. 
Therefore, increasing salt concentration increases the number of
the multivalent counterions surrounding but not condensed on the chain. 
Due to the presence of these surrounding counterions, the chance for a chain 
to escape from its current condensed state is increased. 
The surrounding multivalent counterions strongly interact with the chain,
and can induce local change of the chain morphology by the  
replacing process: some of the condensed ions on the chain 
are released into the solution and replaced with the surrounding ions.
Occasionally, the chain forms a newly condensed structure.

\subsection{Schematic State diagram} 
\label{Sec:state_diagram}
The conformation of our single polyelectrolytes, 
classified by the three categories: 
\textit{flexible} chain,  \textit{semiflexible} chain, and \textit{rigid} chain,
is summarized in a schematic state diagram shown in Fig.~\ref{fig:state_diagram}.

Flexible chain exhibits an extended structure in a salt-free solution, 
and continuously collapse to a compact, spherical structure 
as $C_{s}$ is increased to the equivalence point.   
Further increasing $C_{s}$ gradually leads chain discondensation.  
At very high salt concentration, the chain behaves similar to a random coil. 
 
Semiflexible chain shows different conformations as multivalent salt is 
added into the solution. 
At very low salt concentration, the chain exhibits an elongated structure.
If $C_{s}$ is increased beyond the first critical value $C_c$,
the chain bends over upon itself and forms two-folded rod structure (hairpin).
This structure is not so stable that the chain morphology alters 
between a hairpin and an extended structure.
Further increasing $C_s$ firstly stabilizes the hairpin structure,  
and eventually, leads to the appearance of toroidal structure.
At this moment, hairpin and toroid structures coexist in the solution.
We found that the rate of transition between hairpin and toroid increases 
with the salt concentration.
Therefore, the transition becomes observable during limited time period of a 
molecular simulation while $C_s$ is large enough.
If $C_s$ is not large enough, the two structures can be observed only in
independent runs, since our model system contains solely one chain.
The formation of these ordered structures has been proposed to be 
a consequence of a nucleation-growth process derived from the thermal 
fluctuations in chain morphology~\cite{yoshikawa96b,sakaue02,hud05}.
At higher salt concentration, the chain can even transit between three morphological 
states: the unfolded-rod state, the hairpin state, and the toroid state.
Finally, if $C_{s}$  surpasses the second critical value $C_d$, the chain
can no longer be collapsed, and consequently, reattains a unfolded, rod-like structure. 

Rigid chain, just as its name denotes, is rigid, no matter how many salts 
are added in the solution, and always shows a rodlike, fully stretched structure 
in simulations.  

The state diagram presented here is in accordance with the results 
obtained by other authors, although the employed models are different.
For example, Ou and Muthukumar investigated the conformation of single
polyelectrolytes in a salt-free solution and the chains were collapsed
by increasing Coulomb strength parameter $\Gamma$~\cite{ou05}.
They observed that a semiflexible chain stays, in turn, in a coil
state, a folded-rod state, and a toroid state, 
as $\Gamma$ is increased (cf.~Fig.~6 of Ref.~\cite{ou05}).
The result is consistent with what we observed in the 
region where chains are collapsed.
For the second example, we mention the works done by
Ivanov and his coworkers~\cite{ivanov00,stukan03}.
In their works, the effects of chain stiffness and chain length 
on the conformation of neutral polymers were studied. 
They found that finite chain length can make unsharpened the coil-to-toroid transition 
driven by temperature reduction, and several intermediate states, such as folded-rod and 
disk-like structures, were observed in the transition region. 
The proposed state diagram is in analogy of our results in the low-salt region.
We remark that chain length is an important factor in determination
of the chain morphology. 
Experiments have shown that short DNA molecules are more easily condensed
to folded rods than toroids~\cite{marquet87,arscott90,bloomfield91}.  
This phenomenon gives an explanation of why the hairpin structure appears 
frequently in our state diagram.
It is because the modeled chain is not long in our study.
Nonetheless, the schematic state diagram obtained here has gone beyond 
the works done by other authors.
We investigated the salt-induced conformation of polyelectrolytes 
thoroughly in a \textit{direct} way since this topic has attracted much 
attention in scientific community~\cite{tripathy02,findeis01},
and explored the situations from a salt-free condition to a very high 
salt concentration so that chain condensation and  redissolution are both occurred. 
The findings give deep insight of the phenomena of DNA condensation 
at molecular level and are relevant for a broad range of salt-induced
complexation phenomena.

\section{Conclusion} 
\label{Sec:Conclusion}
We have performed Langevin dynamics simulations to study single polyelectrolytes
in tetravalent salt solutions under the framework of a coarse-grained model. 
The effect of salt concentration and the role of chain stiffness on the properties of 
chain conformation have been discussed. In order to understand thoroughly the reentrant
condensation, a broad range of salt concentration has been investigated so that 
the single polyelectrolytes display, in turn, collapsing transition and
swelling transition upon addition of salt.

We have observed that chain morphology crucially depends on chain stiffness. 
At a fixed chain length, the radius of gyration $R_{g}$ of 
a polyelectrolyte varies in three different ways with the salt concentration $C_{s}$.
Based upon these variations, we can classify polyelectrolytes into three categories.  
The first category is \textit{flexible} chain for which $\left<R_g^2\right>^{1/2}$
shows a tilted L-shaped curve as a function of $C_{s}$ in the  semilog plot.  
The second one is \textit{semiflexible} chain which is characterized by 
an U-shaped $\left<R_g^2\right>^{1/2}$ curve.
And the third one is \textit{rigid} chain for which $\left<R_g^2\right>^{1/2}$
is roughly a constant, no matter how many salts are added in the solution. 
The end-to-end distance shows the similar variation as the radius of gyration.
However, its value does not always reflect the real chain size, 
particularly when the chain is condensed to an orderly packed structure such as hairpin. 
Moreover, we have found that the WLC model with persistence length obtained by the OSF
theory qualitatively predicts the end-to-end distance at low and at high salt concentrations  
for semiflexible and rigid chains.
Surprisingly, these theories also give a not-bad description of the decreasing 
behavior of $\left<R_e^2\right>^{1/2}$ against $C_s$ for flexible chain up to $C_s=C_s^*$, 
although they are not expected to work well neither for flexible chains  
nor for a solution with multivalent salt.

We have found that in a salt-free or vary low-salt solution, the degree of swelling 
of a polyelectrolyte decreases with increasing the chain stiffness, 
in reference of the size of the uncharged counterpart of the chain. 
On the other hand, in a high-salt region, it tends toward zero, regardless of 
the chain stiffness.  As a consequence, at high $C_{s}$, a semiflexible 
chain reattains a dimension close to the chain size in the absence of salt, 
but not does a flexible one.
We have discussed the reasons for the occurrence of chain reentrance or swelling 
at high salt concentration.  For flexible chains, two effects can possibly 
drive this phenomenon: (1) the net Coulomb repulsion between condensed 
multivalent counterions which locally overcharges the chain, (2) the
instability derived from the presence of non-condensed multivalent 
counterions in a bulk solution. For semiflexible chains, the chain 
stiffness plays, in addition, the imperative role on reexpanding the chain. 

The chain stiffness has been witnessed to influence both the local and the global 
structures of a polyelectrolyte.
A flexible chain shows a zigzagged local structure in the presence of multivalent 
salt and the condensed structure is disordered.
In contrast, a semiflexible chain locally shows smooth morphology.
Large-angle turn on the chain backbone can only be observed  when
the chain is condensed to a hairpin structure. 
Moreover, the global structure of a condensed chain is orderly packed. 
Hairpin and toroid are the two favored structures.

The chain stiffness affects also the nature of the structural transition.
By studying the probability density distributions of asphericity $A$ and of $R_g$,
we have demonstrated that a semiflexible chain undergoes a coil-globule transition 
in a discrete manner between an extended structure and a condensed, ordered structure, 
whereas a flexible chain undergoes a continuous transition.
This discrete nature is happened not only in the low-salt region when
chain condensation occurs, but also in the high-salt region when chain 
reenters into the solution.

We have studied the dynamics of the conformational quantities of 
a semiflexible chain.  Upon addition of salt beyond the first critical 
concentration $C_{c}$, the chain is folded to a hairpin structure. 
However, it is a dynamic equilibrium process and the chain morphology 
alters between a hairpin and a unfolded-chain structure. 
Further increasing $C_{s}$ firstly stabilizes the hairpin structure and 
then triggers the birth of toroid structure. 
Transition between hairpin and toroid becomes more frequent as $C_s$ getting higher. 
When $C_{s}$ approaches the second critical concentration $C_d$, 
these two structures become unstable and
the chain can stay again in a unfolded state, coexisting with the two condensed 
structures.  
At the end, we summarized the conformation of individual 
polyelectrolytes in a state diagram.  
Although the modeled chain is not very long, our results agree with experiments and 
other numerical works in many aspects in the chain condensation region.
The work presented here goes further to investigate in detail the microstructure
and the dynamics of a chain in the chain redissolution region, which 
gives us a thorough vision of the behavior of polyelectrolytes in salt solutions. 
Since chain length is also a relevant factor to affect the morphology of a chain, 
it will be very interesting to study the effect of finite chain length. 
That will be the topic of our future work. 

\section{Acknowledgments} 
This material is based upon work supported by the National Science Council,
the Republic of China, under the contract No.~NSC 95-2112-M-007-025-MY2.
A large part of the simulations was run using the resources of the National Center 
for High-performance Computing under the project 
``Taiwan Knowledge Innovation National Grid''. 
The authors express their gratitude to the members and the staffs of
the council and the center.


\newpage
\section*{Figure captions}
\begin{itemize}
\item[FIG 1] 
$\left<R_g^2\right>^{1/2}$ as a function of $C_s$
for the chains of different stiffness $k_{a}$.

\item[FIG 2]
$\alpha\equiv (\left<R_g^2\right>^{1/2}/\left<R_{g,n}^2\right>^{1/2})-1 $
as a function of $C_s$ for the chains of different stiffness $k_a$.

\item[FIG 3]
$\alpha $ as a function of $k_a$ at different salt
concentration $C_s$.

\item[FIG 4]
$\left< R_e^2\right>^{1/2}$ as a function of $C_s$ for
the chains of different $k_{a}$.
The dashed curves denote the prediction of the WLC model
combined with the OSF and MO theories
for different values of $k_{a}$, indicated on the left-hand side of the
corresponding curve.  

\item[FIG 5]
Snapshots of simulation for a flexible chain ($k_a=0k_BT/rad^2$)
at (a) $C_s=1.33\times 10^{-5}\sigma^{-3}$,
(b) $C_s=8\times 10^{-5}\sigma^{-3}$,
and (c) $C_s=5.12\times 10^{-3}\sigma^{-3}$.
The gray bead-spring chain represents the polyelectrolyte,
the black spheres represent the tetravalent counterions,
and the white spheres represent the monovalent counterions.
Coions are not shown for the reason of clarity.

\item[FIG 6]
Snapshots of simulation for a semiflexible chain ($k_a=8k_BT/rad^2$)
at (a) $C_s=1.33\times 10^{-5}\sigma^{-3}$,
(b) $C_s=6.4\times 10^{-4}\sigma^{-3}$,
and (c) $C_s=5.12\times 10^{-3}\sigma^{-3}$.
The gray bead-spring chain represents the polyelectrolyte,
the black spheres represent the tetravalent counterions,
and the white spheres represent the monovalent counterions.
Coions are not shown for the reason of clarity.

\item[FIG 7]
Snapshots of simulation for a rigid chain ($k_a=100k_BT/rad^2$)
at $C_s=8\times 10^{-5}\sigma^{-3}$.
The gray bead-spring chain represents the polyelectrolyte,
the black spheres represent the tetravalent counterions,
and the white spheres represent the monovalent counterions.
Coions are not shown for the reason of clarity.

\item[FIG 8]
(a) $p(A)$, and (b) $p(R_g)$ of the flexible chain
with zero stiffness for the cases of different $C_s$.
Each case is numbered in Roman number along `Case'-axis.
The value in the parenthesis following the Roman number denotes the
salt concentration in the unit $10^{-5}\sigma^{-3}$.   

\item[FIG 9]
(a) $p(A)$, and (b) $p(R_g)$ of the semiflexible chain
with $k_{a}=8k_B T/rad^2$ for the cases of  different $C_s$.
Each case is numbered in Roman number along `Case'-axis.
The value in the parenthesis following the Roman number denotes the
salt concentration in the unit $10^{-5}\sigma^{-3}$. 

\item[FIG 10]
(a) $p(A)$, and (b) $p(R_g)$ of a rigid chain with $k_{a}=100k_B T/rad^2$
for the cases of different $C_s$.
Each case is numbered in Roman number along `Case'-axis.
The value in the parenthesis following the Roman number denotes the
salt concentration in the unit $10^{-5}\sigma^{-3}$.

\item[FIG 11]
Time evolution of (a) $A$,  and (b) $R_g$ and $R_e$,
at $C_s= 8\times 10^{-5} \sigma_m^{-3}$ where
the chain exhibits a hairpin structure.

\item[FIG 12]
Time evolution of (a)  $A$, and (b) $R_g$ and $R_e$,
at $C_s= 8\times 10^{-5} \sigma_m^{-3}$ where
the  chain shows a toroidal structure.

\item[FIG 13]
Time evolution of (a) $A$, and (b) $R_g$ and $R_e$,
at $C_s= 6.4\times 10^{-4} \sigma_m^{-3}$ where
the chain alters between a hairpin and a toroid structure. 

\item[FIG 14]
Time evolution of (a) $A$, and (b) $R_g$ and $R_e$,
at $C_s= 1.28\times 10^{-3} \sigma_m^{-3}$ where the chain 
morphology alters between a hairpin, a toroid, and an unfolded-chain
structure.

\item[FIG 15]
Schematic state diagram for our single polyelectrolytes.
A double-headed arrow appeared between two states denotes that
the structural transition can take place in a typical simulation run at that condition.

\end{itemize}

\pdfoutput=1 

\newpage
\pagestyle{empty}

\newpage
\pagestyle{empty}

\newpage
\pagestyle{empty}

\newpage
\pagestyle{empty}

\newpage
\pagestyle{empty}

\newpage
\pagestyle{empty}

\newpage
\pagestyle{empty}

\newpage
\pagestyle{empty}

\newpage
\pagestyle{empty}
\setcounter{figure}{7} 
\begin{figure} 
\begin{center}
\includegraphics[height=\textwidth,angle=90]{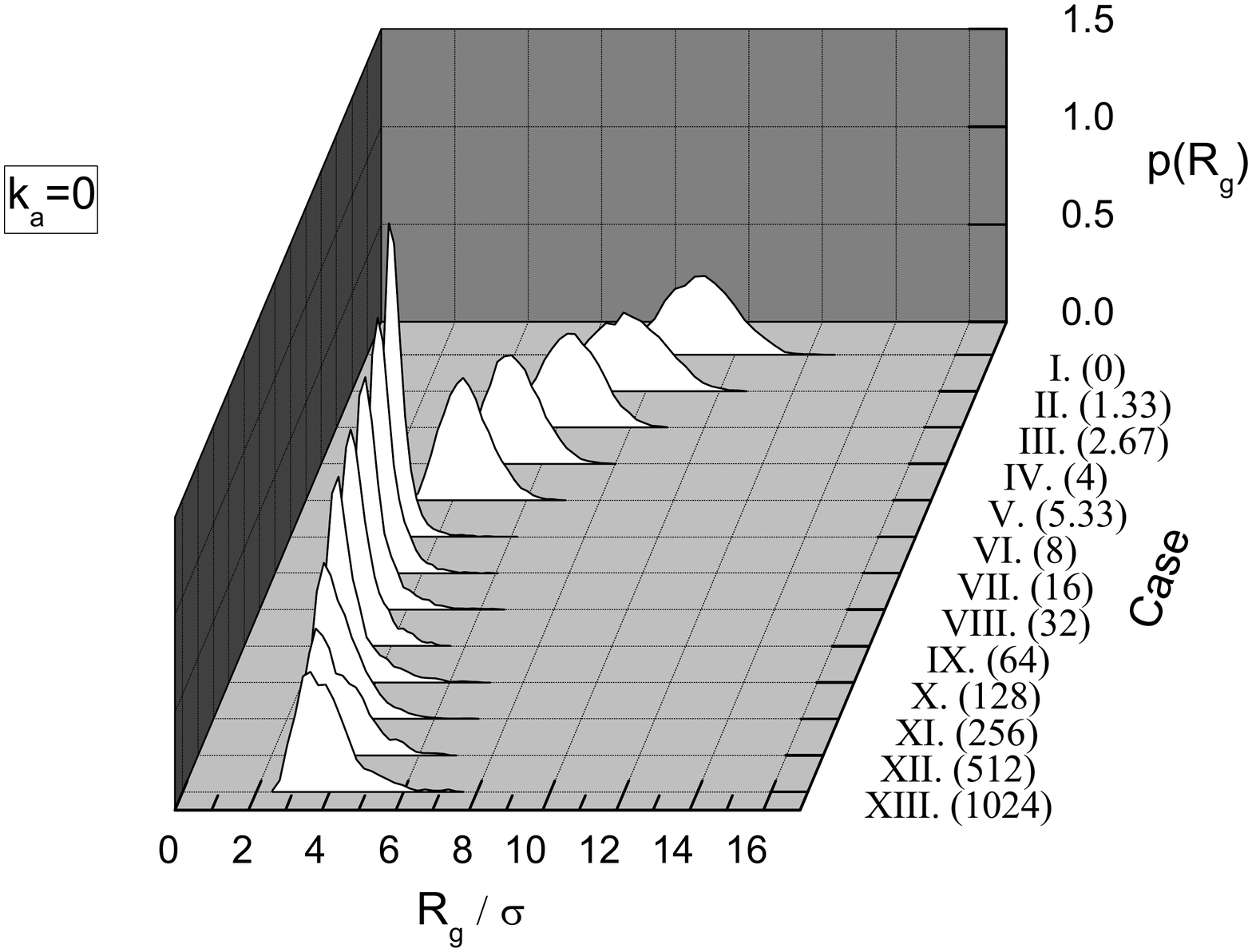}
\caption{(b)}
\end{center}
\end{figure} 

\newpage
\pagestyle{empty}
\begin{figure}
\begin{center}
\includegraphics[height=\textwidth,angle=90]{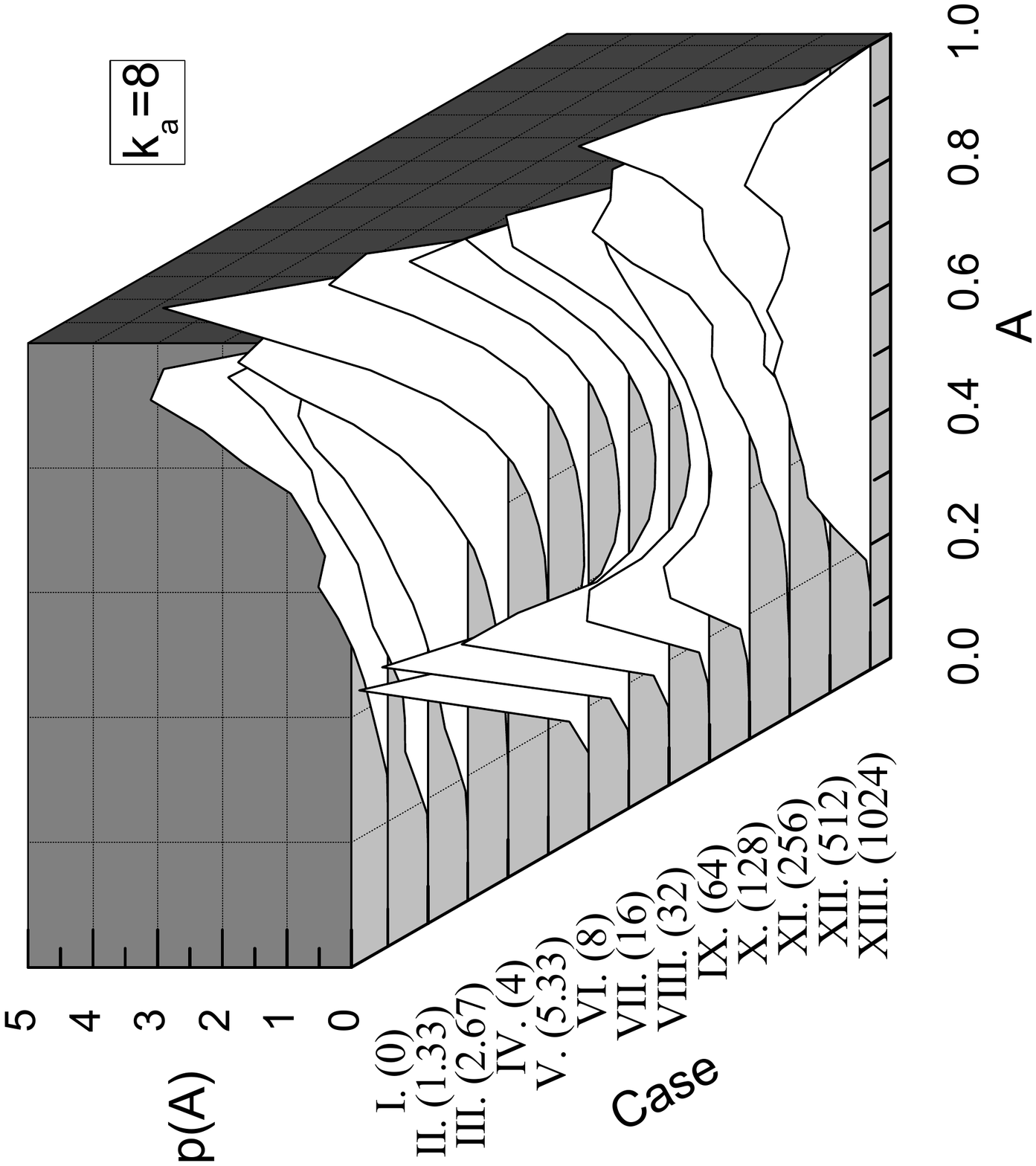}
\caption{(a)}
\label{fig:histoD_ka8}
\end{center}
\end{figure} 

\newpage
\pagestyle{empty}
\setcounter{figure}{8} 
\begin{figure} 
\begin{center}
\includegraphics[height=\textwidth,angle=90]{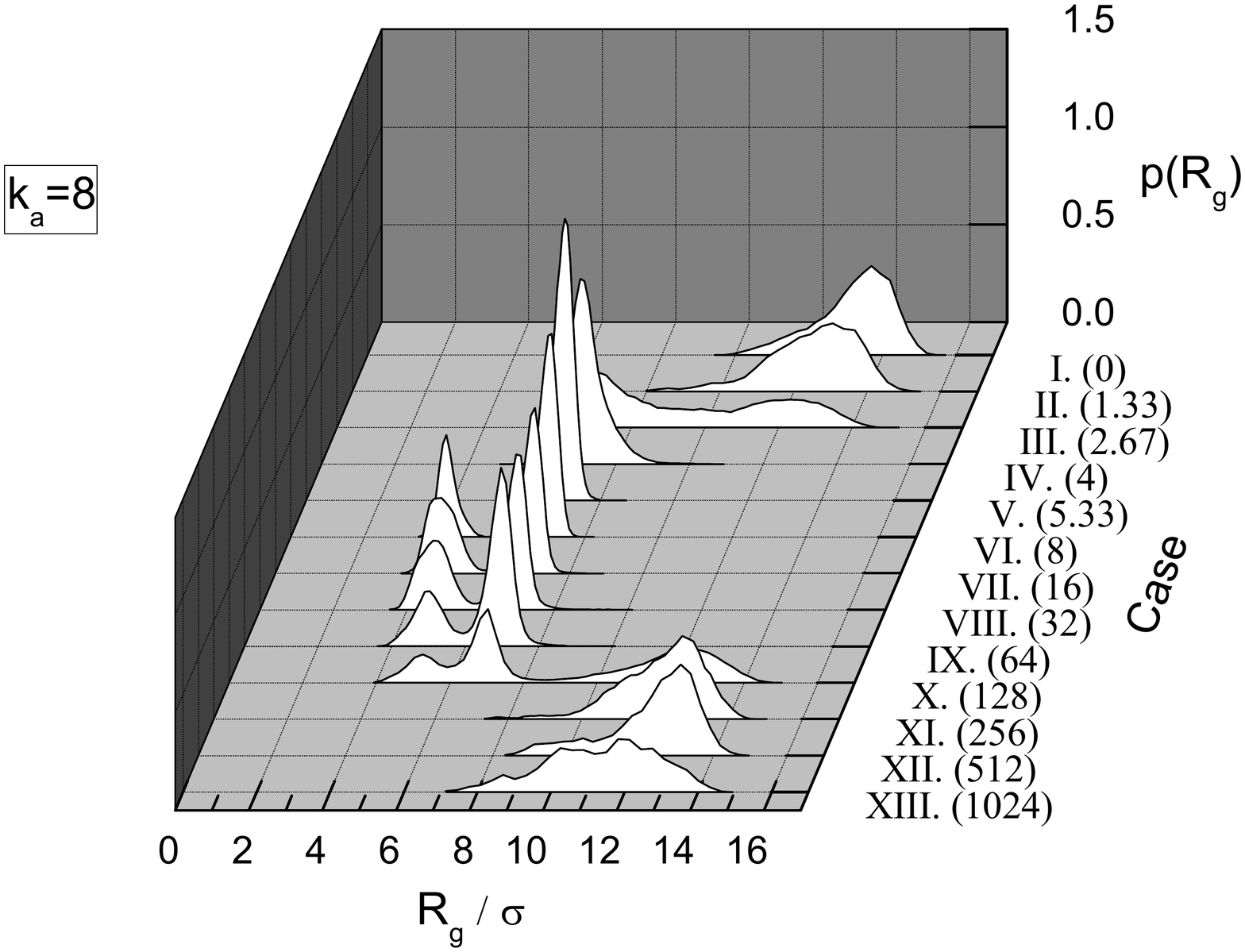}
\caption{(b)}
\end{center}
\end{figure} 

\newpage
\pagestyle{empty}
\begin{figure} 
\begin{center}
\includegraphics[height=\textwidth,angle=90]{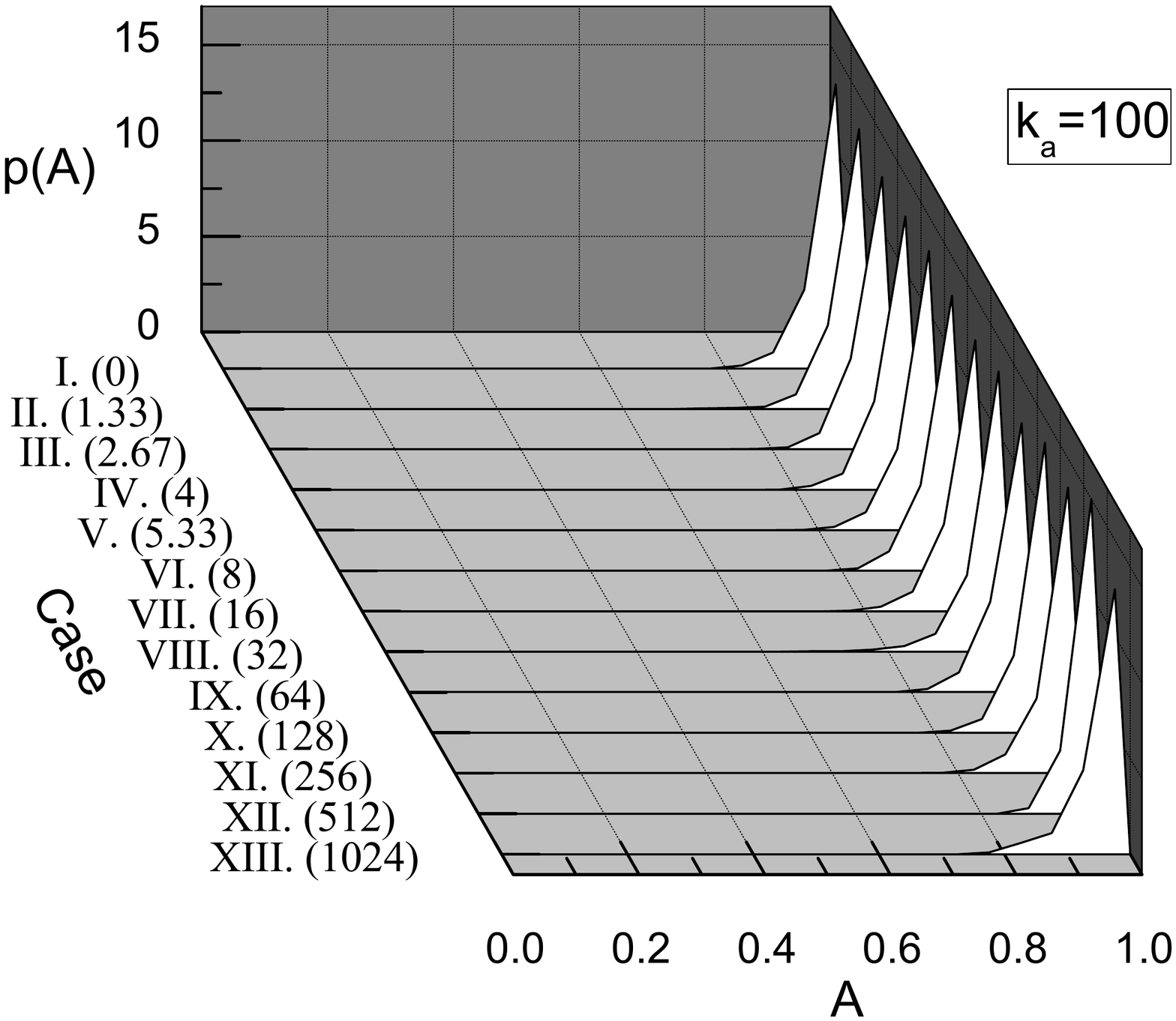}
\caption{(a)}
\label{fig:histoD_ka100}
\end{center}
\end{figure} 

\newpage
\pagestyle{empty}
\setcounter{figure}{9} 
\begin{figure} 
\begin{center}
\includegraphics[height=\textwidth,angle=90]{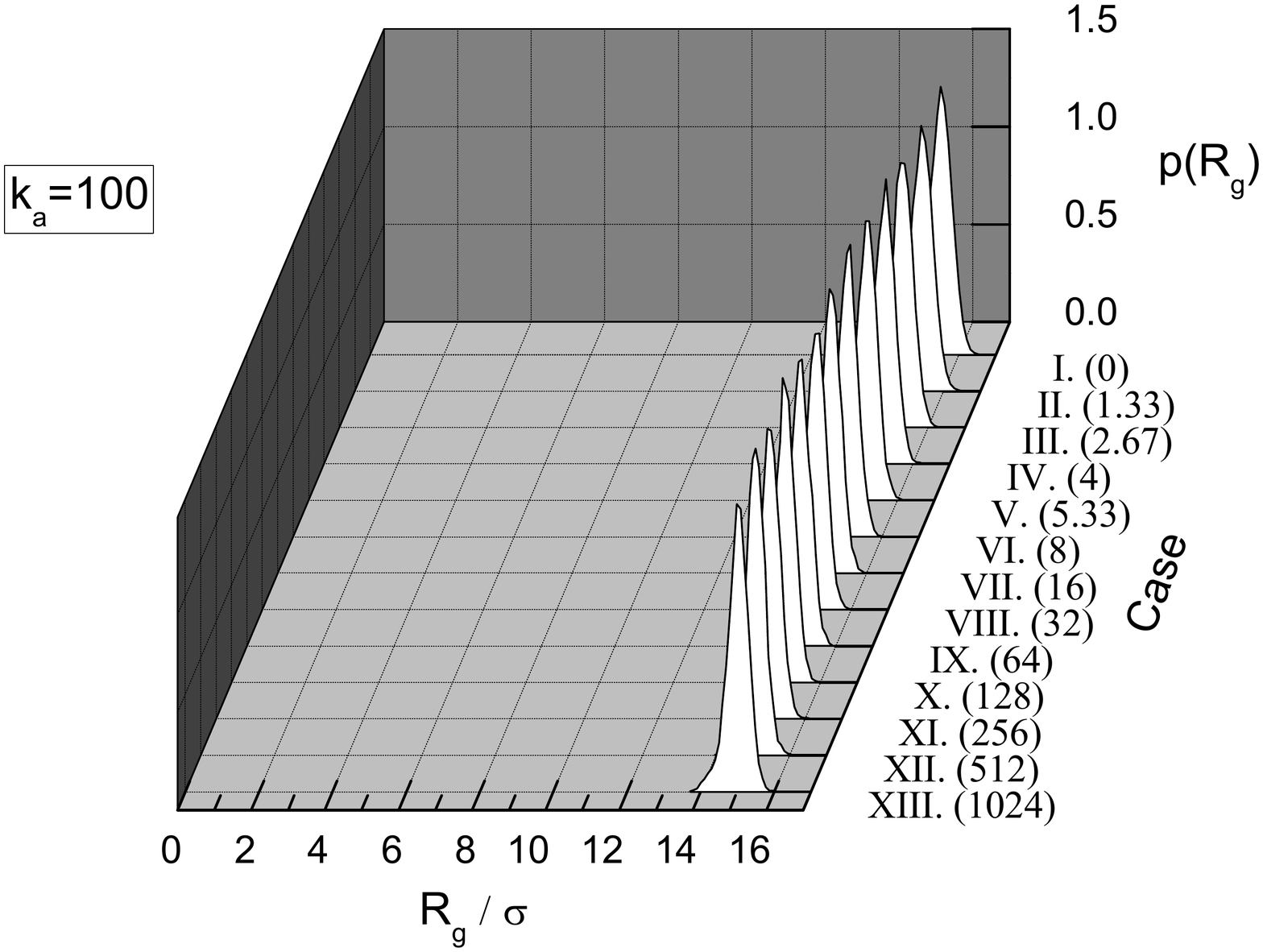}
\caption{(b)}
\end{center}
\end{figure} 

\newpage
\pagestyle{empty}
\begin{figure} 
\begin{center}
\includegraphics[height=\textwidth,angle=90]{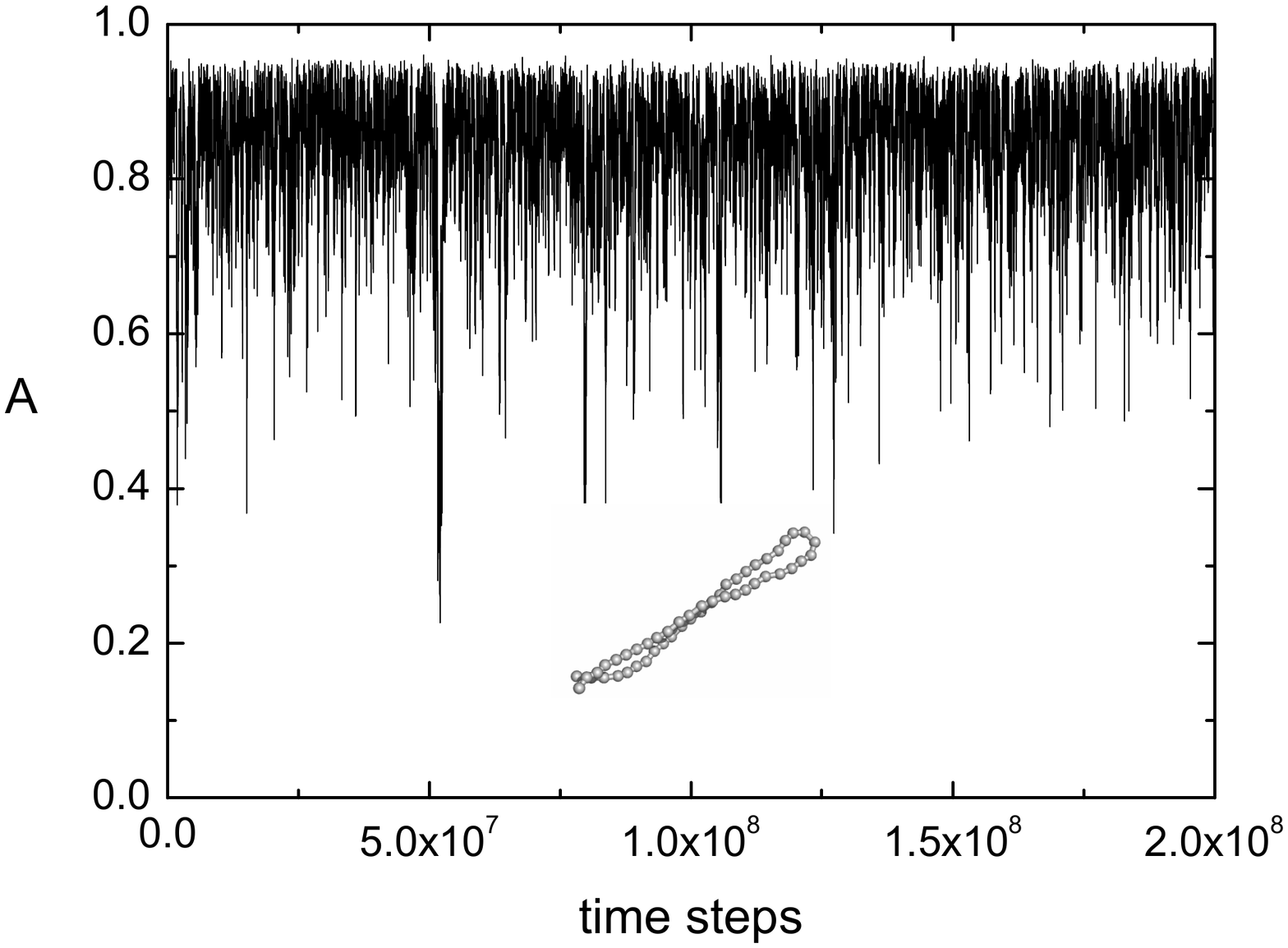}
\caption{(a)}
\label{fig:yt_plot_S8c4}
\end{center}
\end{figure} 

\newpage
\pagestyle{empty}
\setcounter{figure}{10} 
\begin{figure} 
\begin{center}
\includegraphics[height=\textwidth,angle=90]{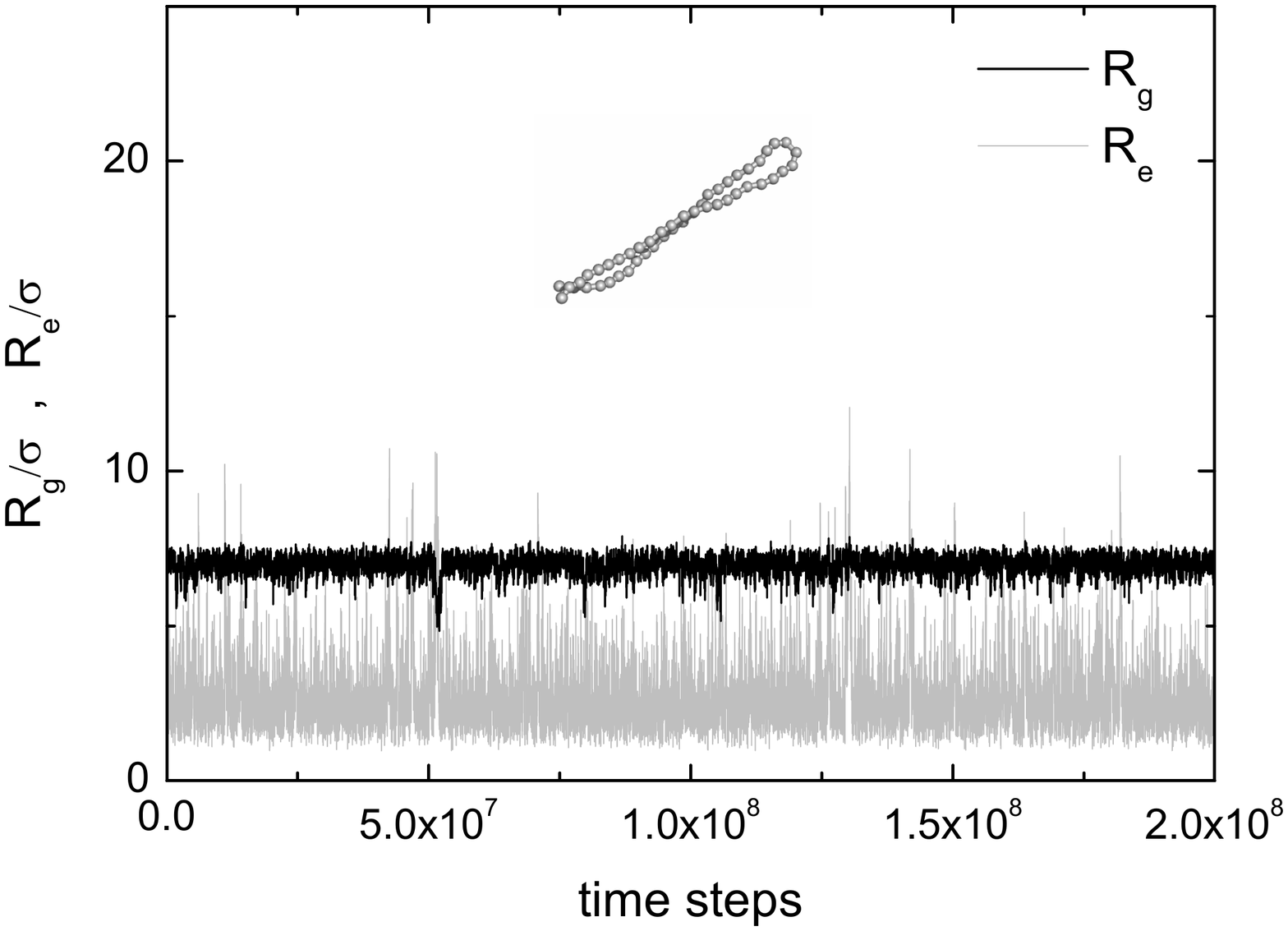}
\caption{(b)}
\end{center}
\end{figure} 

\section*{Figures 1---15}

\newpage
\pagestyle{empty}
\begin{figure} 
\begin{center}
\includegraphics[height=\textwidth,angle=90]{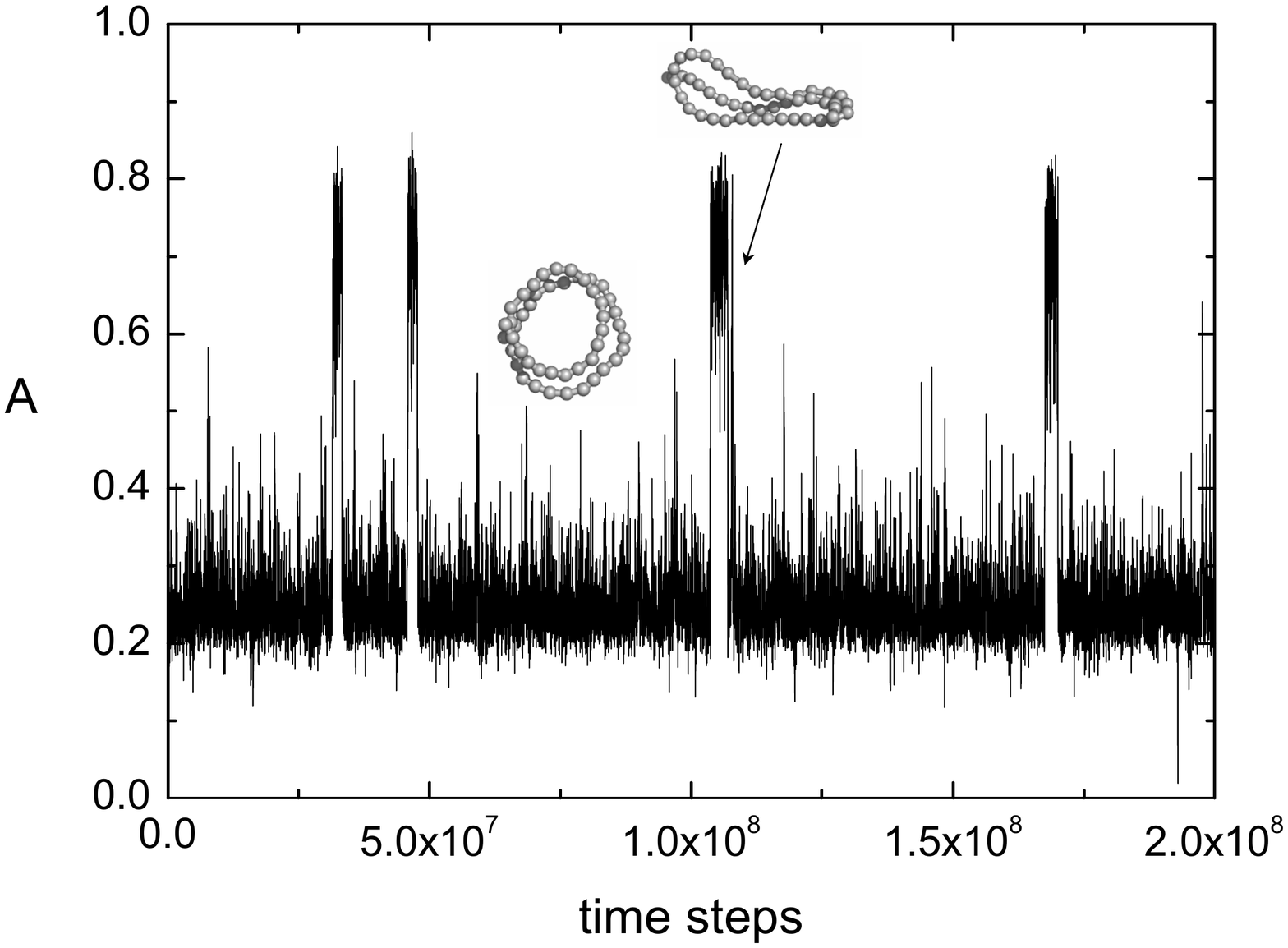}
\caption{(a)}
\label{fig:yt_plot_S8c5}
\end{center}
\end{figure} 

\newpage
\pagestyle{empty}
\setcounter{figure}{11} 
\begin{figure}
\begin{center}
\includegraphics[height=\textwidth,angle=90]{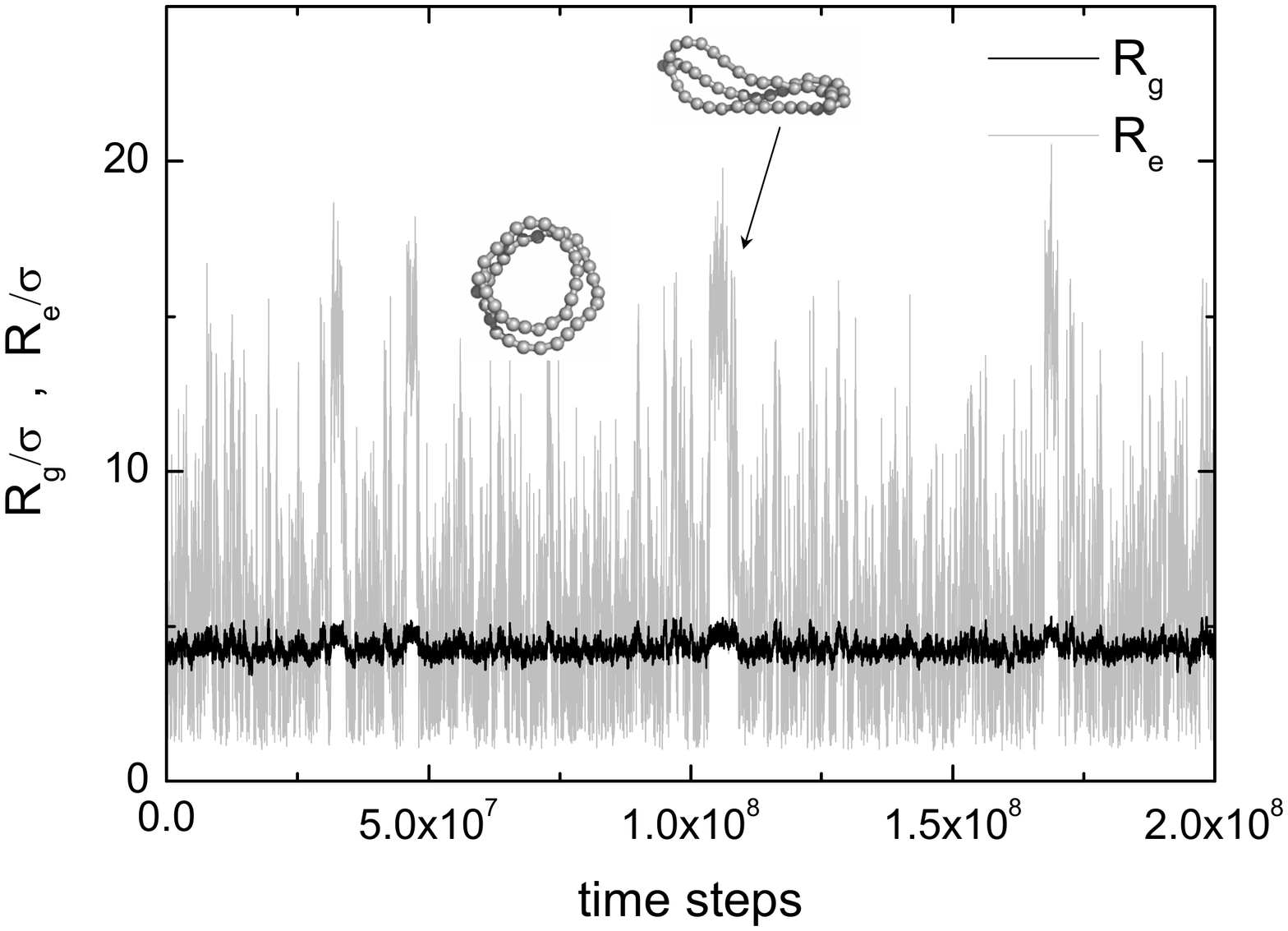}
\caption{(b)}
\end{center}
\end{figure} 

\newpage
\pagestyle{empty}
\begin{figure}
\begin{center}
\includegraphics[height=\textwidth,angle=90]{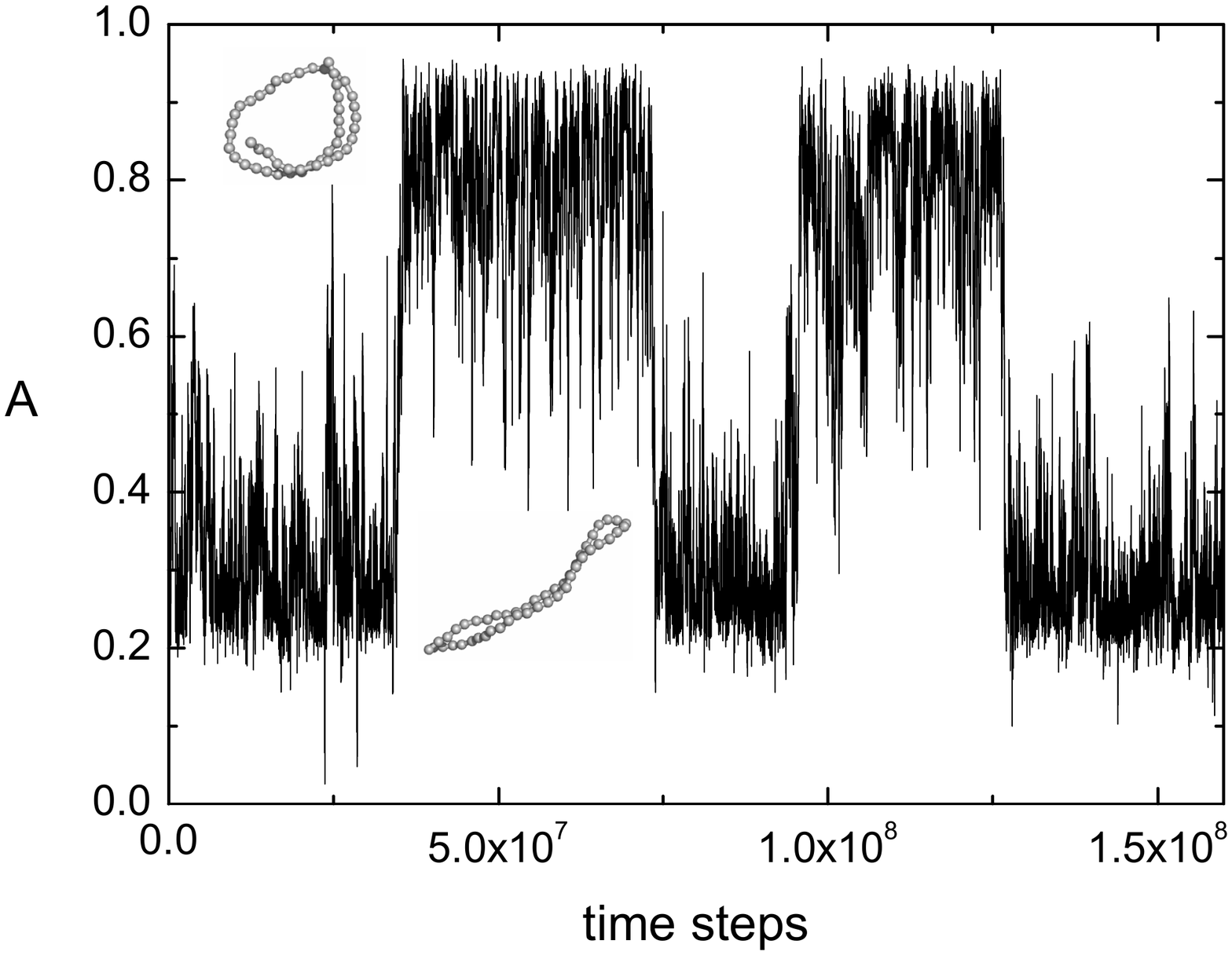}
\caption{(a)}
\label{fig:yt_plot_S96}
\end{center}
\end{figure} 

\newpage
\pagestyle{empty}
\setcounter{figure}{12} 
\begin{figure}
\begin{center}
\includegraphics[height=\textwidth,angle=90]{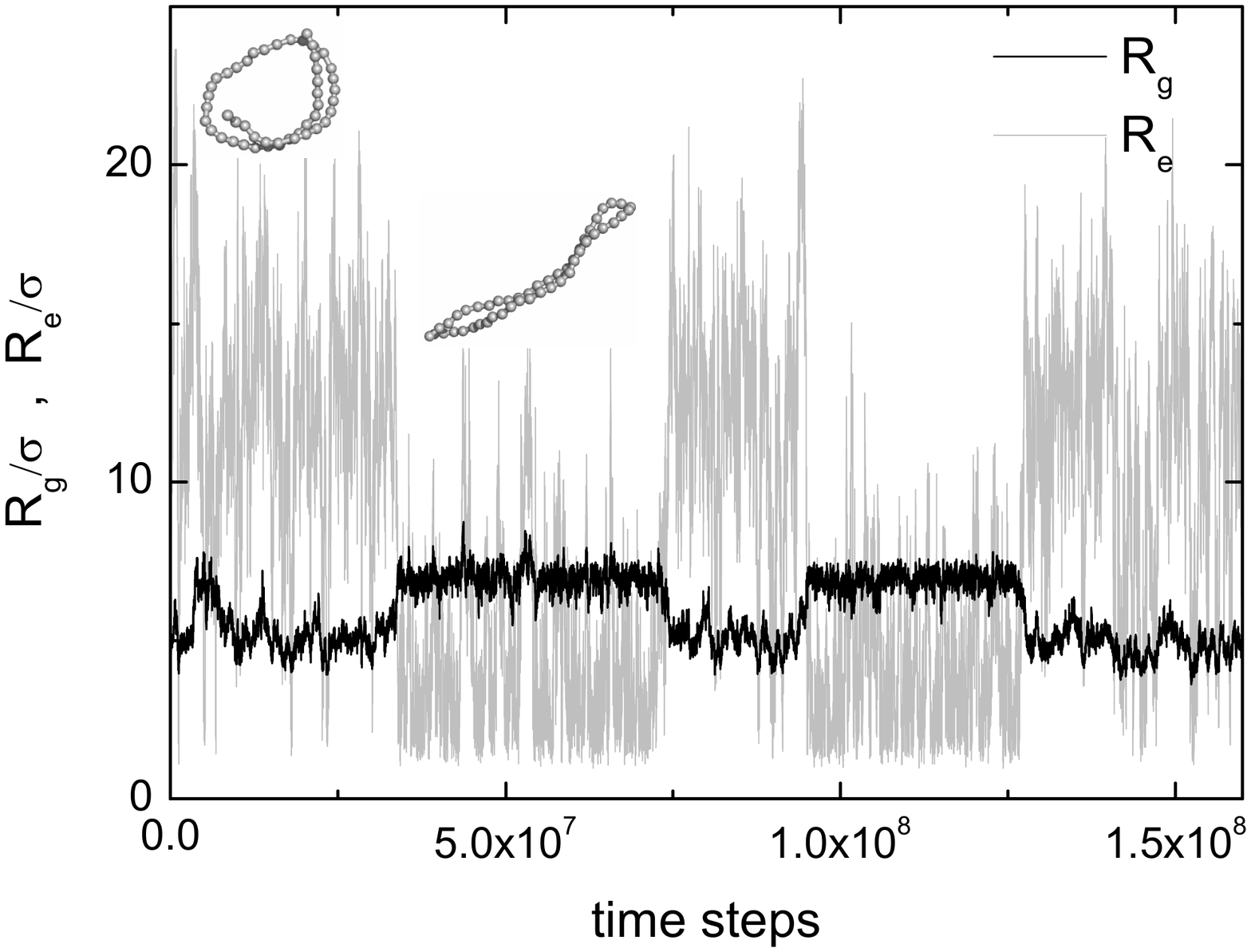}
\caption{(b)}
\end{center}
\end{figure} 

\newpage
\pagestyle{empty}
\begin{figure}
\begin{center}
\includegraphics[height=\textwidth,angle=90]{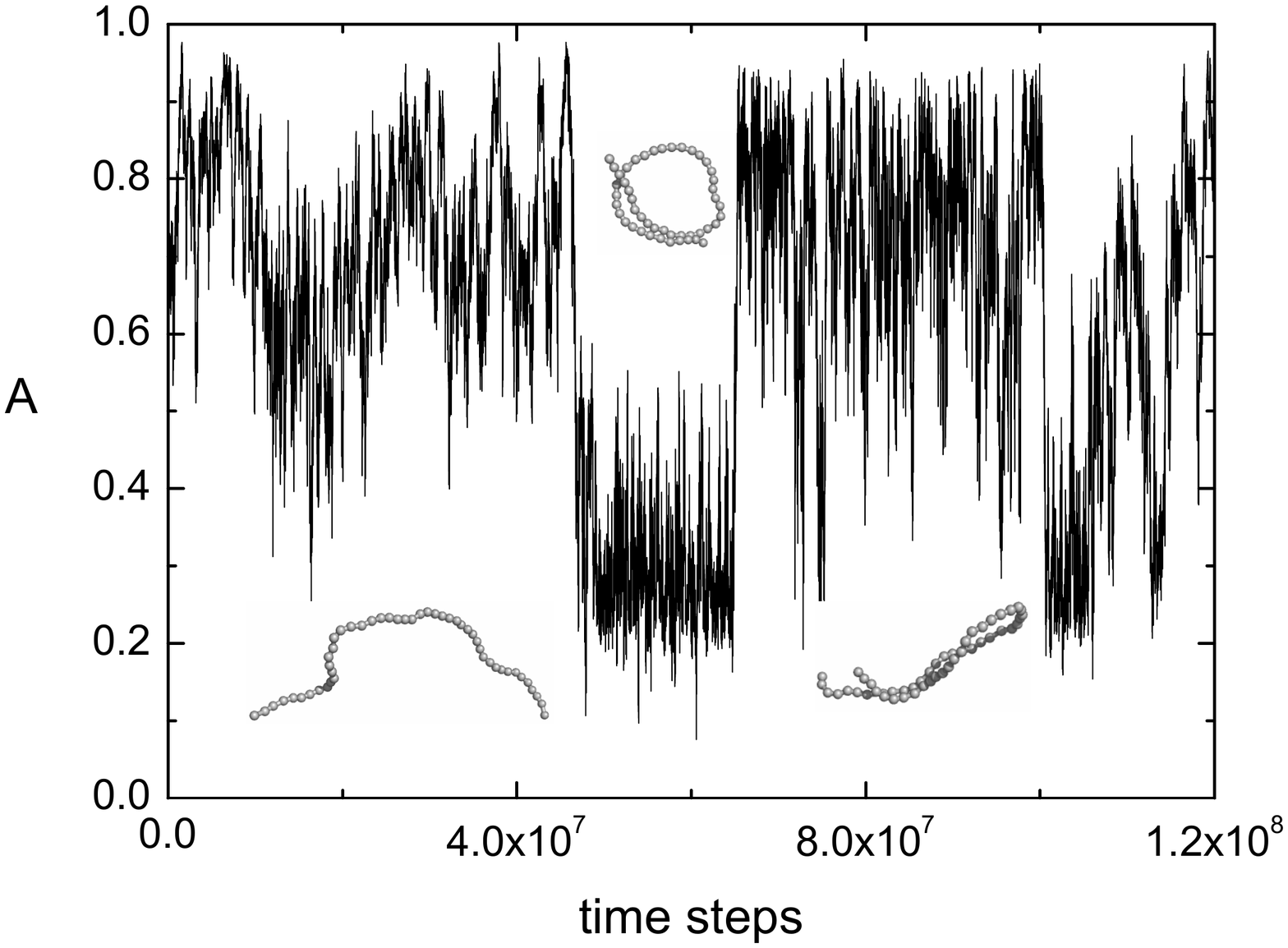}
\caption{(a)}
\label{fig:yt_plot_S192}
\end{center}
\end{figure} 

\newpage
\pagestyle{empty}
\setcounter{figure}{13} 
\begin{figure}
\begin{center}
\includegraphics[height=\textwidth,angle=90]{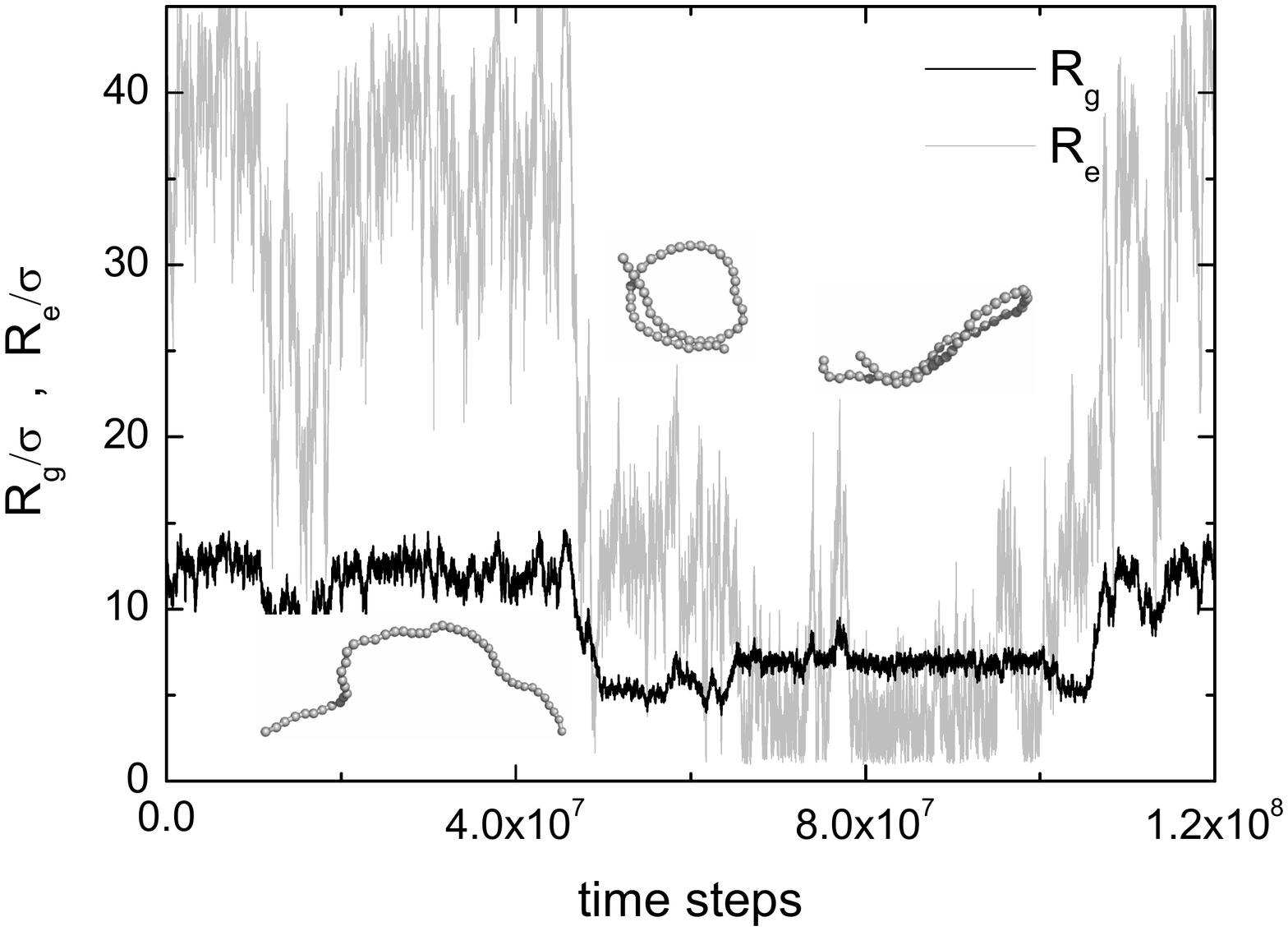}
\caption{(b)}
\end{center}
\end{figure} 

\newpage
\pagestyle{empty}
\begin{figure}
\begin{center}
\includegraphics[height=\textwidth,angle=90]{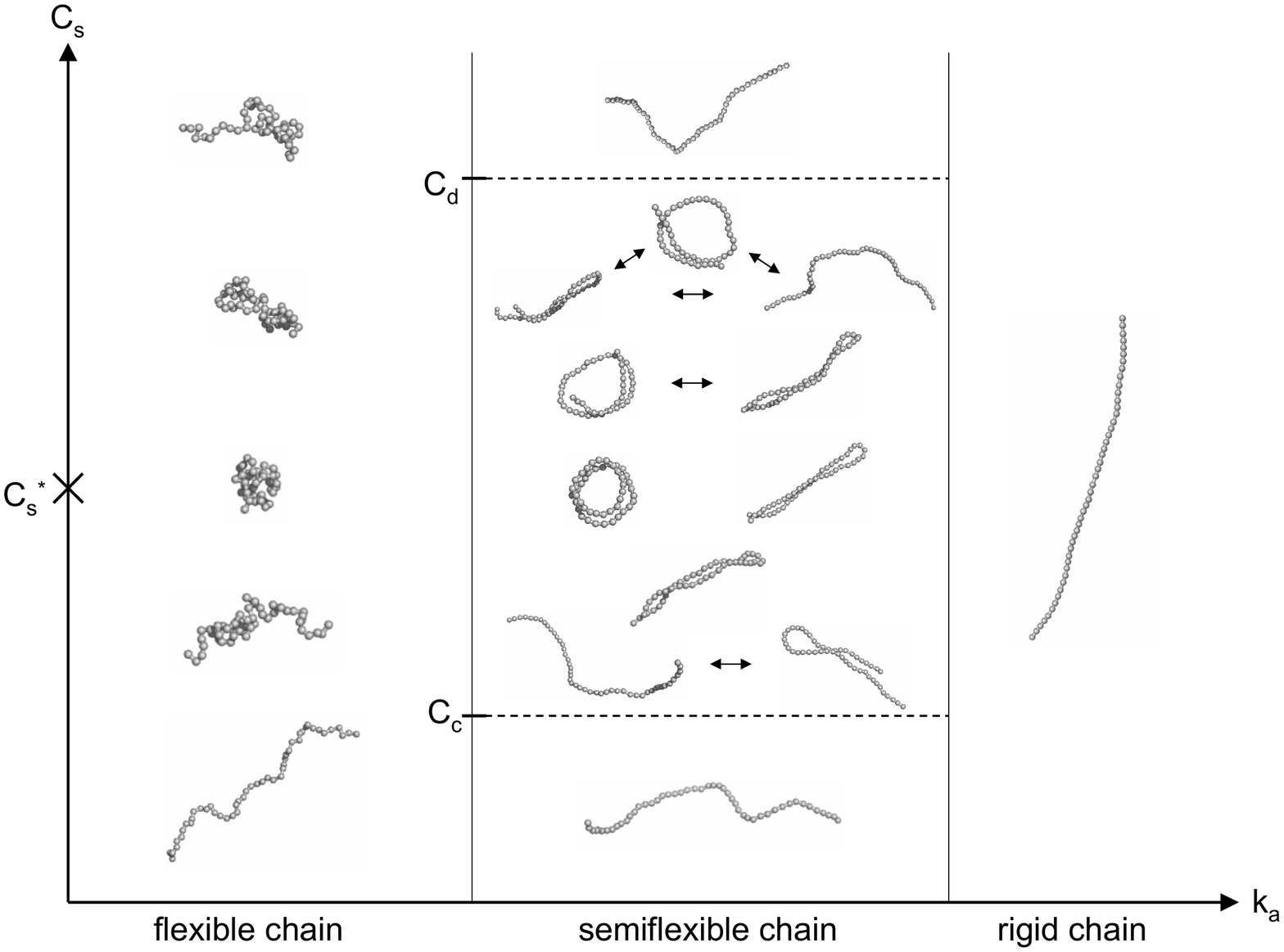}
\caption{}
\label{fig:state_diagram}
\end{center}
\end{figure} 

\end{document}